\documentclass[twocolumn]{aastex7}

\usepackage{float}
\usepackage{booktabs}
\usepackage{rotating}

\newcommand{\photoz}{photo-$z$}
\newcommand{\Photoz}{Photo-$z$}
\newcommand{\photozs}{photo-$z$'s}

\newcommand{\zlp}{$z_\mathrm{LePHARE}$}
\newcommand{\specz}{spec-$z$}

\newcommand{\speczs}{spec-$z$'s}

\newcommand{\zspeczero}{$z_\mathrm{spec}=0.0$}

\newcommand{\nmad}{$\sigma_\mathrm{NMAD}$}
\newcommand{\fout}{$f_\mathrm{outlier}$}

\newcommand{\lephare}{\texttt{LePHARE}}
\newcommand{\lepharez}{\texttt{LePHARE}-$z$}
\newcommand{\lepharezs}{\texttt{LePHARE}-$z$'s}
\newcommand{\umapz}{UMAP-$k$NN-$z$}
\newcommand{\somz}{SOM-$z$}
\newcommand{\colorsz}{colors-$k$NN-$z$}
\newcommand{\nearestz}{nearest-$z$}
\newcommand{\nneighbors}{\texttt{n\_neighbors}}
\newcommand{\mindist}{\texttt{min\_dist}}

\newcommand{\secref}[1]{Section~\ref{sec:#1}}
\newcommand{\appref}[1]{Appendix~\ref{app:#1}}
\newcommand{\tabref}[1]{Table~\ref{tab:#1}}

\newcommand{\figref}[1]{Figure~\ref{fig:#1}}
\newcommand{\figrefs}[2]{Figures~\ref{fig:#1} and \ref{fig:#2}}

\begin{document}

\title{Optimizing Photometric Redshift Training Sets I: Efficient Compression of the Galaxy Color--Redshift Relation with UMAP}

\correspondingauthor{Finian Ashmead}
\email{finianashmead@pitt.edu}

\author[0000-0001-9416-5874]{Finian Ashmead}
\affiliation{Department of Physics and Astronomy, University of Pittsburgh, Pittsburgh, PA 15260, USA}
\affiliation{Pittsburgh Particle Physics, Astrophysics, and Cosmology Center (PITT PACC), University of Pittsburgh, Pittsburgh, PA 15260, USA}
\email{finianashmead@pitt.edu}

\author[0000-0001-8684-2222]{Jeffrey A. Newman}
\affiliation{Department of Physics and Astronomy, University of Pittsburgh, Pittsburgh, PA 15260, USA}
\affiliation{Pittsburgh Particle Physics, Astrophysics, and Cosmology Center (PITT PACC), University of Pittsburgh, Pittsburgh, PA 15260, USA}
\email{janewman@pitt.edu}

\author[0000-0001-8085-5890]{Brett H. Andrews}
\affiliation{Department of Physics and Astronomy, University of Pittsburgh, Pittsburgh, PA 15260, USA}
\affiliation{Pittsburgh Particle Physics, Astrophysics, and Cosmology Center (PITT PACC), University of Pittsburgh, Pittsburgh, PA 15260, USA}
\email{andrewsb@pitt.edu}

\author[0000-0001-5063-8254]{Rachel Bezanson}
\affiliation{Department of Physics and Astronomy, University of Pittsburgh, Pittsburgh, PA 15260, USA}
\affiliation{Pittsburgh Particle Physics, Astrophysics, and Cosmology Center (PITT PACC), University of Pittsburgh, Pittsburgh, PA 15260, USA}
\email{rachel.bezanson@pitt.edu}

\author[0000-0002-5665-7912]{Biprateep Dey}
\affiliation{Department of Statistical Sciences, University of Toronto, Toronto, ON M5G 1Z5, Canada}
\affiliation{Canadian Institute for Theoretical Astrophysics (CITA), University of Toronto, Toronto, ON M5S 3H8 }
\affiliation{Dunlap Institute for Astronomy \& Astrophysics, University of Toronto, Toronto, ON M5S 3H4, Canada}
\affiliation{Vector Institute, Toronto, ON M5G 0C6, Canada}
\email{biprateep@pitt.edu}


\author[0000-0001-5382-6138]{Daniel C. Masters}
\email{dmasters@ipac.caltech.edu}
\affiliation{Caltech/IPAC, 1200 E. California Blvd. Pasadena, CA 91125, USA}

\author[0000-0003-0122-0841]{S.A. Stanford}
\affiliation{Physics and Astronomy Department,
University of California
Davis, CA 95616}
\email{stanford@physics.ucdavis.edu}

\begin{abstract}

Spectroscopic datasets are essential for training and calibrating photometric redshift (\photoz) methods.  However, spectroscopic redshifts (\speczs) constitute a biased and sparse sampling of the photometric galaxy population, which creates difficulties for the common grid-based approach for mapping color to redshift using self-organizing maps (SOMs).  Instead, we utilized the uniform manifold approximation and projection (UMAP) algorithm to compress a Rubin--Roman-like $ugrizyJH$ color space into a thin and densely-sampled manifold.  Crucially, the manifold varies continuously and monotonically in redshift and specific star formation rate in roughly orthogonal directions.  \added{Using COSMOS2020 many-band \photozs\ and compiled \speczs\ } as representative and non-representative samples, respectively, we trained and tested redshift \added{prediction} from a SOM, from nearest neighbors in UMAP coordinates (\umapz)\added{, and directly from nearest neighbors in the color space to assess how well location in each space maps to redshift.  }
\added{For the representative training set, \umapz\ exhibited smaller \photoz\ scatter and fraction of outliers than SOM-based methods. When training with the highly-biased \specz\ sample, \umapz\ maintained similar performance while SOM- and color-based methods predictions were significantly degraded, especially at $z>1.5$ where training sets are the most sparse.}
The physically-meaningful trends across the UMAP manifold allow for accurate redshift prediction even in 
\added{such poorly-sampled regions of color space.}
This suggests that representative, spectroscopically-anchored training sets can be produced by interpolating between spectroscopic sources at the UMAP coordinates of photometric objects, maximizing the performance of \photoz\ algorithms.

\end{abstract}

\keywords{Redshift --- Cosmology --- Galaxy Evolution --- Machine Learning --- Dimensionality Reduction}

\section{Introduction} \label{sec:intro}


Photometric redshifts (\photozs) are critical to extragalactic science. Cosmological measurements via weak lensing and large-scale structure are particularly dependent on \photozs\ to accurately characterize redshift distributions for large samples of galaxies \citep{Mandelbaum2018}. For upcoming experiments, characterization of \photoz\ distributions is expected to be a leading source of systematic uncertainty (\citealt{DESCSRD}, \citealt{Newman2022}).

Photometric redshift techniques fall broadly into two categories, template-based and machine-learning-based, both of which rely on spectroscopic redshifts (\speczs) in calibration or training data \citep{Salvato2019}. Both classes of techniques are limited by sparse sampling of color space, and the systematic differences in magnitude, color, and redshift distributions between photometric and spectroscopic samples. Machine-learning-based techniques in particular make a strong assumption that the \specz\ training data is representative of the target sample to which the \photoz\ algorithm will be applied, and deviation from this assumption limits their performance \citep{Newman2022}.

Real spectroscopic datasets deviate from being representative samples of larger photometric datasets for several reasons. Spectroscopic surveys are generally designed with cuts on the magnitude, color, or surface brightness of target galaxies in order to ensure high success rates within limited exposure time or target galaxies in a particular redshift range. This introduces selection effects into the spectroscopic dataset that will not reflect the distributions of galaxy properties in more inclusive photometric surveys
. 
Spectroscopic datasets also include unintentional selection effects, resulting from non-uniform probabilities of obtaining a secure redshift measurement for different galaxy populations across redshifts, as well as unavoidable selection effects due to atmospheric absorption in certain wavelength ranges, most notably the infrared. These effects result broadly in higher-redshift objects and dimmer objects across redshifts being under-represented in spectroscopic datasets. For instance, \citet{Khostovan2025} found that spectroscopic observations in the COSMOS field have poor coverage of intermediate-mass quiescent galaxies and low-to-intermediate-mass ``bursty" star-forming galaxies at $z<1$, massive quiescent galaxies at $z\sim2$, and massive star-forming galaxies at $z>3$. 

Deep spectroscopic datasets generally come from a limited area on the sky, with their redshift distributions reflecting the cosmic structure along that particular line of sight, which will not be representative of the general redshift distribution of galaxies. Different populations of galaxies also cluster differently, introducing further bias when these populations are not represented proportionally in the spectroscopic dataset (as is generally the case in practice). Furthermore, the limited sky area of these surveys entails limited volume at low redshift, making it difficult to obtain representative samples in this regime. 
Finally, even high-confidence spectroscopic redshift datasets contain a non-negligible fraction of incorrect redshifts, degrading \photoz\ performance.

Photometric redshift techniques are fundamentally mappings of observed galaxy spectral energy distributions (SEDs) to redshift. There are many possible degeneracies in particular bands or colors, for example the Balmer/$4000$Å break at $z\sim0.3$ and the Lyman break at $z\sim3$ would both be observed as a steep rise in flux between the \textit{g} and \textit{r} bands. These degeneracies can often be lifted with the inclusion of measurements in additional bands, with the tradeoff that increasing the dimensionality of the color space exponentially increases its volume, leading to sparse sampling by the data \citep{McCullough2024}. 
This is known as the curse of dimensionality, and makes interpolating in high-dimensional spaces difficult and unreliable. 

Observed galaxy colors however, are highly correlated and occupy only a subset of the space, being largely driven by only two parameters, redshift and specific star formation rate (sSFR).
With a proper application of dimensionality reduction, it should be possible to recover this structure. 
Self-organizing maps (SOMs; \citealt{Kohonen1982})
have become a popular approach 
to embed higher-dimensional galaxy photometry into a two-dimensional grid and map the cells to redshift, sSFR, and other physical galaxy properties (e.g., \citealt{Carrasco2014}, \citealt{Masters2015}, \citealt{Hemmati2019}, \citealt{Khostovan2025}). 

The SOM algorithm assigns objects to cells on a regular grid, \added{which loses some information to binning} while also inhibiting it from capturing intrinsic structure in the data that may be better represented using a different geometry. A SOM can be used to assign redshifts to objects based on the known redshifts of sources in the same cell. However, major discontinuities in the redshifts of objects occupying adjacent cells make it infeasible to interpolate reliably across the map for objects in cells without any known redshifts.
Here, we explore alleviating these drawbacks using the uniform manifold approximation and projection algorithm (UMAP; \citealt{McInnes2018}), which has the advantage of projecting all of the data into a \textit{continuous} low-dimensional space and is designed to capture both local and global structure in datasets. 
By predicting redshifts from the low-dimensional coordinates and quantifying the performance with common \photoz\ evaluation metrics, we quantify the reliability of the color--redshift mapping in these compressed spaces. 
\added{By comparing to redshift prediction in the input color space itself, we also quantify the tradeoffs between the information loss inherent to compression (losing 4-5 dimensions) and the benefits of working in a lower-dimensional space which is more densely sampled by training data.}

In Section \ref{sec:data}, we describe the data used in this analysis. In Section \ref{sec:methods}, we detail our creation of SOM and UMAP embeddings of our input data, and describe the resulting spaces qualitatively. In Section \ref{sec:redshift_estimation}, we describe our quantitative assessment of the color--redshift mappings provided by the SOM and UMAP embeddings, comprising their use as redshift estimators and the metrics used to evaluate their performance. In Section \ref{sec:discussion}, we comment on these results and the future direction of this work.

\section{Data}
\label{sec:data}

In our analysis we used photometry and template-based, many-band photometric estimates of redshift and sSFR from the Farmer version of the COSMOS2020 catalog \citep{Weaver2022, https://doi.org/10.26131/irsa564} and spectroscopic redshifts from the COSMOS Spectroscopic Redshift Compilation (\citealt{Khostovan2025}; hereafter CSRC). \added{The COSMOS field features extensive multiwavelength imaging along with the most complete and least biased spectroscopy of any sizeable field on the sky, making it a standard testing area for \photoz\ research. 
For this reason, it has been selected as a deep drilling field with intensive observations by the Vera C. Rubin Observatory, the Nancy Grace Roman Space Telescope, and other surveys (\citealt{LSST2025}, \citealt{ROTAC}, \citealt{Treyer2026}, \citealt{Giannini2026}). The $ugrizyJH$ photometry in the COSMOS2020 catalog covers comparable wavelength range and depths to what will be available in the combined Rubin and Roman catalogs.}

\subsection{COSMOS2020}
\label{sec:COSMOS2020}
\added{In this work we use photometry as well as many-band \photozs\ and sSFRs from COSMOS2020, which constituted the most complete and accurate set of \photozs\ available when it was undertaken, with the best coverage of the COSMOS area (the COSMOS2025 catalog incorporates more data enabling better \photozs, but only over a limited area; see \citealt{Shuntov2025}). Our primary motivation is to improve \photozs\ for the Rubin and Roman lensing samples. COSMOS2020 spans the range of galaxy properties that would be included in these samples while providing the closest available approximation of ideal \specz\ coverage (i.e., with a selection which is independent of both color and magnitude). The COSMOS2020 \photozs\ and CSRC redshift compilation (\secref{CSRC}) are the most comprehensive current datasets that approach LSST gold sample depth, and are expected to dominate the training and calibration of Rubin and Roman photometric redshifts in the near future.}

For the COSMOS2020 catalog, we removed areas of the field with partial coverage or unreliable photometry, which we implemented by requiring \texttt{FLAG\_COMBINED}=0, reducing the sample from 964,506 sources to 746,976. We corrected for Milky Way extinction according to the dust map of \citet{SFD1998}. We also corrected for the systematic photometric offsets derived from \lephare\ based on COSMOS spectroscopic sources, following the COSMOS2020 example Jupyter notebook \citep{cosmos2020readcat}.

The COSMOS2020 catalog features estimates of galaxy properties (e.g., redshift, sSFR) calculated using two different SED template-fitting codes, \texttt{EAZY} \citep{Brammer2008} and \lephare\ (\citealt{Arnouts1999}, \citealt{Ilbert2006}). \lephare\ finds the combination of template and redshift that minimizes the chi-squared with respect to the observed photometry,
and in the case of COSMOS2020 was implemented in the configuration used by \citet{Ibert2013}. \texttt{EAZY} is implemented similarly, with differences in the template library and fitting method used. We adopted the \lephare\ redshift estimates as they had a lower outlier rate than \texttt{EAZY} when evaluating with \speczs\ (using the metrics defined in \secref{performance_metrics}; \added{see \citealt{Yin2025} for another application of similar metrics}).

The \lephare\ many-band photometric redshifts in COSMOS2020 (\lepharezs) have the advantage of being based on up to $30$ broad, medium, and narrow bands spanning the UV to IR wavelengths (with as few as $17$ bands being used for objects in our sample), which provides more accurate redshift estimates than broadband-only \photozs\ can. 
\added{For our quantitative analysis, we require redshift labels for a sample of galaxies representative of those from imaging datasets like those that will be produced by Rubin and Roman; COSMOS2020 \photoz\ estimates are available for all objects with photometry in the catalog, fulfilling this need.}

\added{We will compare these labels to redshift estimates based only on $ugrizyJH$ colors, including algorithms trained with the highly-biased spectroscopic sample. By testing each technique using a fixed set of COSMOS2020 redshifts, we can quantify the differences in redshift prediction performance between dimensionality reduction methods as well as the impact of training with the highly-biased spectroscopic sample as compared to a fully-representative random sample of \lepharezs\ of the same size.}

\added{The \lepharezs\ are the most reliable redshift estimates available for many galaxies and hence have been used to supplement \speczs\ in redshift calibration for cosmological experiments (for example, in \citealt{Yin2025}, \citealt{Giannini2026}, \citealt{Nishizawa2020}, and \citealt{Treyer2026}).}
\added{However, }they do have a somewhat larger outlier rate compared to some spectroscopic datasets, particularly \added{for the very faintest galaxies}, with rates approaching 4\% for objects with $24.25<i<24.5$. 
\added{The metric we use to assess scatter between \lepharezs\ and predicted redshifts, the normalized median absolute deviation (\nmad), is highly robust to outliers and hence will not be impacted by such a rate of incorrect \lepharezs\ (see Section \ref{sec:performance_metrics} for definitions of the metrics used).  For the outlier rate (\fout) and bias, any causes of catastrophic outlier redshifts in the \lepharezs\ would provide an advantage to the \lephare-trained estimates as compared to the \specz-trained estimates, as the former should share the same failure modes but the latter would not. This is in addition to any advantages from the training sample being representative of the redshift and color distributions of the target sample when training with \lepharezs, which would not hold for the \specz-trained case.}

We applied three additional cuts to the COSMOS2020 catalog. We removed sources with \lephare\ redshifts \zlp\ $<0.01$ due to their significant likelihood of being misclassified stars, bringing the dataset down to 709,024 sources. 
We also restricted our analysis to sources with both $i<24.5$ and $H<24.8$, comparable to the expected magnitude limits for early LSST or Roman weak lensing analyses, leaving us with 121,791. These cuts also helped to eliminate objects that may have large photometric uncertainties and potentially poorly-constrained photometric redshifts. 

Throughout this paper, we applied dimensionality reduction methods to a seven-dimensional parameter space consisting of the $u-g$, $g-r$, $r-i$, $i-z$, $z-y$, $y-J$, and $J-H$ colors of the objects. This approximates the colors between successive bands that will be available by combining the medium tier of the Roman High-Latitude Wide Area Survey (HLWAS) with the Rubin Legacy Survey of Space and Time (LSST) \citep{ROTAC}.
We utilized the combined $u$-band photometry presented in COSMOS2020 (\citealt{Sawicki2019}, \citealt{Weaver2022}). The $grizy$ data in COSMOS2020 were taken from the second data release of the Hyper Suprime-Cam Subaru Strategic Program \citep{Aihara2019}. The $J$ and $H$ near-infrared data in COSMOS2020 were taken from the fourth data release of the UltraVISTA survey \citep{McCracken2012}.
We tested alternate sets of eight colors (adding $J-K_s$) and ten colors (adding the Spitzer IRAC-based colors $K_s-Ch1$ and $Ch1-Ch2$), obtaining results similar or worse to those based on the seven-color space.

In addition to the cuts described above, we removed all sources missing measurements in any of the photometric bands defining the input color space, or \lephare\ \photoz\ or sSFR. 
\added{We also removed 22 sources with \zspeczero\ and \zlp\ $>0.0$ (i.e., stars misidentified by \lephare).}
In the end, our sample includes \added{112,387} sources. 

\subsection{COSMOS Spectroscopic Redshift Compilation}
\label{sec:CSRC}
The COSMOS Spectroscopic Redshift Compilation (CSRC; \citealt{Khostovan2025}) compiled $\sim$165,000 \speczs\ of $\sim$98,000 unique objects taken in the COSMOS field over the course of $\sim$20 years in 108 observing programs. 
To utilize CSRC \speczs\, we performed a spatial cross-match to our COSMOS2020 photometric data, 
considering the nearest COSMOS2020 source (with a maximum separation of $0.5^{\prime\prime}$) as the counterpart of a CSRC object. 
We obtained 35,346 matches, with a large fraction of unmatched CSRC objects being in regions outside of the formal COSMOS2020 catalog area or in masked regions within it.

The CSRC compiled redshifts from observing programs using various quality flagging systems, so the authors normalized confidence levels (CL) and quality flags ($Q_f$) based on the popular $0-4$ scheme for all CSRC redshifts to create a consistent scale. From our cross-matches we selected sources with CL $>95$, corresponding to $Q_f=4$ or a \textit{very reliable redshift}, obtaining a sample of 14,499 objects.
\added{Of these objects, 22 had \zspeczero\, which we have removed as probable stars (as mentioned above). This leaves 14,477 objects,}
hereafter referred to as the \specz\ CL $>95$ sample. 

\begin{figure*}
\centering
\includegraphics[width=\textwidth]{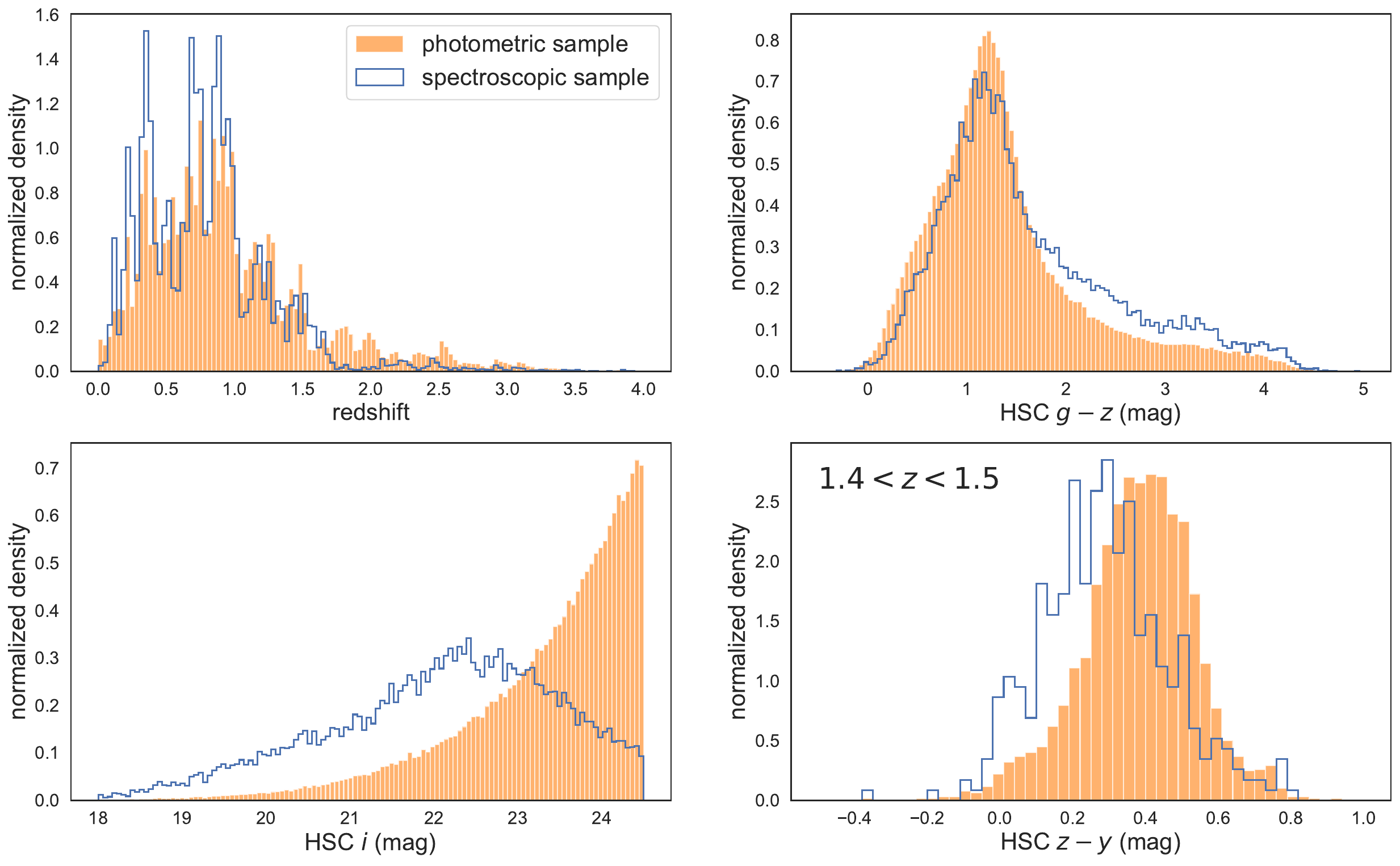}
\caption{\label{fig:non_IID} The distributions of our photometric (orange) and spectroscopic (\specz\ CL $>95$; blue) samples in redshift (upper left), $g-z$ color (all redshifts; upper right), $i$-band magnitude (lower left), and $z-y$ color for objects at redshifts $1.4<z<1.5$ (lower right). Higher redshifts, particularly at $z>1.5$, are underrepresented in the spectroscopic sample, as are objects fainter than $i\sim23$. The $g-z$ color distribution is redder overall for the spectroscopic sample, while the $z-y$ color at $1.4<z<1.5$ is systematically bluer for the spectroscopic sample. In the case of the $g-z$ distribution, the apparent overabundance of spectroscopic sources at $g-z\gtrsim1.5$ reflects redder galaxies at $z<1$, while in the case of the $z-y$ distribution for $1.4<z<1.5$, the bluer colors of the spectroscopic sample reflect a preference for star forming galaxies. 
Offset color distributions at fixed redshift such as that shown in the lower right will bias color--redshift mappings based on non-representative spectroscopic samples.
The spectroscopic sample's biased coverage of color space can also be seen in the bottom row of \figref{som_50_100}.}
\end{figure*}

In \figref{non_IID}, we show density-normalized histograms illustrating the differences in the distributions in redshift, color, and magnitude between the spectroscopic and photometric samples. It is notable that even in this case where the observations target the same field, and therefore trace the same line of sight through cosmic structure, the spectroscopically-observed objects are still clearly not representative of the larger photometric sample. For data taken from different fields the existing discrepancies would be much starker, particularly in the case of the fluctuations in redshift distributions.
\added{This will be the case for near-future studies with Rubin and Roman, where training data will come from relatively small areas within deep drilling fields. }

\section{Dimensionality Reduction Methods} \label{sec:methods}

SOM-based approaches to mapping the high-dimensional space of galaxy colors to physical properties and redshift are quite popular, but are limited by the discrete grid and enforced topology such that interpolation across cells is not robust. 
\added{Furthermore, binning the galaxies into discrete cells constitutes an additional compression and loss of some information contained in the structure of the data in the input space. }
By using UMAP as an alternative approach to compressing the high-dimensional space of galaxy colors, we seek to overcome these limitations that hinder redshift prediction.

\subsection{SOM} \label{sec:methods_som}

We implemented SOM using the \texttt{minisom} Python package \citep{minisom}. For a given dataset, one of the most critical choices is that of the SOM size; a higher cell count will capture the structure of the input space in greater detail, but the cells will be increasingly sparsely populated. The overall shape of the SOM and its constituent cells can also be customized (see \citealt{Carrasco2014}), but for the purposes of this investigation we adopt the convention of using rectangular SOM embeddings with a $1\times2$ aspect ratio consisting of rectangular cells as in, e.g., \citet{Masters2015}. 

In our testing of the accuracy of redshift predictions based on position in SOM or UMAP coordinates, we split our dataset into training, \added{validation,} and test samples, and \added{predicted} the redshifts of the test sources based on the training sources, comparing to the \lephare\ \photoz\ for evaluation. 
\added{The validation sample was used for optimizing the hyperparameters of all algorithms and for estimating the uncertainties associated with that process.  It is constructed to have equal size to the \specz\ CL $>95$ sample but contains none of the same galaxies.}


We generated rectangular SOM embeddings for various total cell counts between 128 ($8\times16$) and 11,250 ($75\times150$)\added{, for values of the learning rate hyperparameter between 0.2 and 1.0, and for values of the sigma hyperparameter between 0.5 and 4. Distances in the color space were computed with either the Euclidean (i.e., square root of the sum of squares) or Manhattan (i.e., sum of absolute values) distance metric. Quantitative tests involved in optimizing these choices were run using the validation dataset. }

\added{Our best results were obtained with a $75\times150$ SOM, generated using the Manhattan distance metric with a learning rate of 0.8 and a sigma of 1. 
These choices are explained in greater detail in \appref{som_optimize}. To train to convergence, we used two million learning epochs.}

\begin{figure}
    \centering
    \includegraphics[width=\linewidth]
    {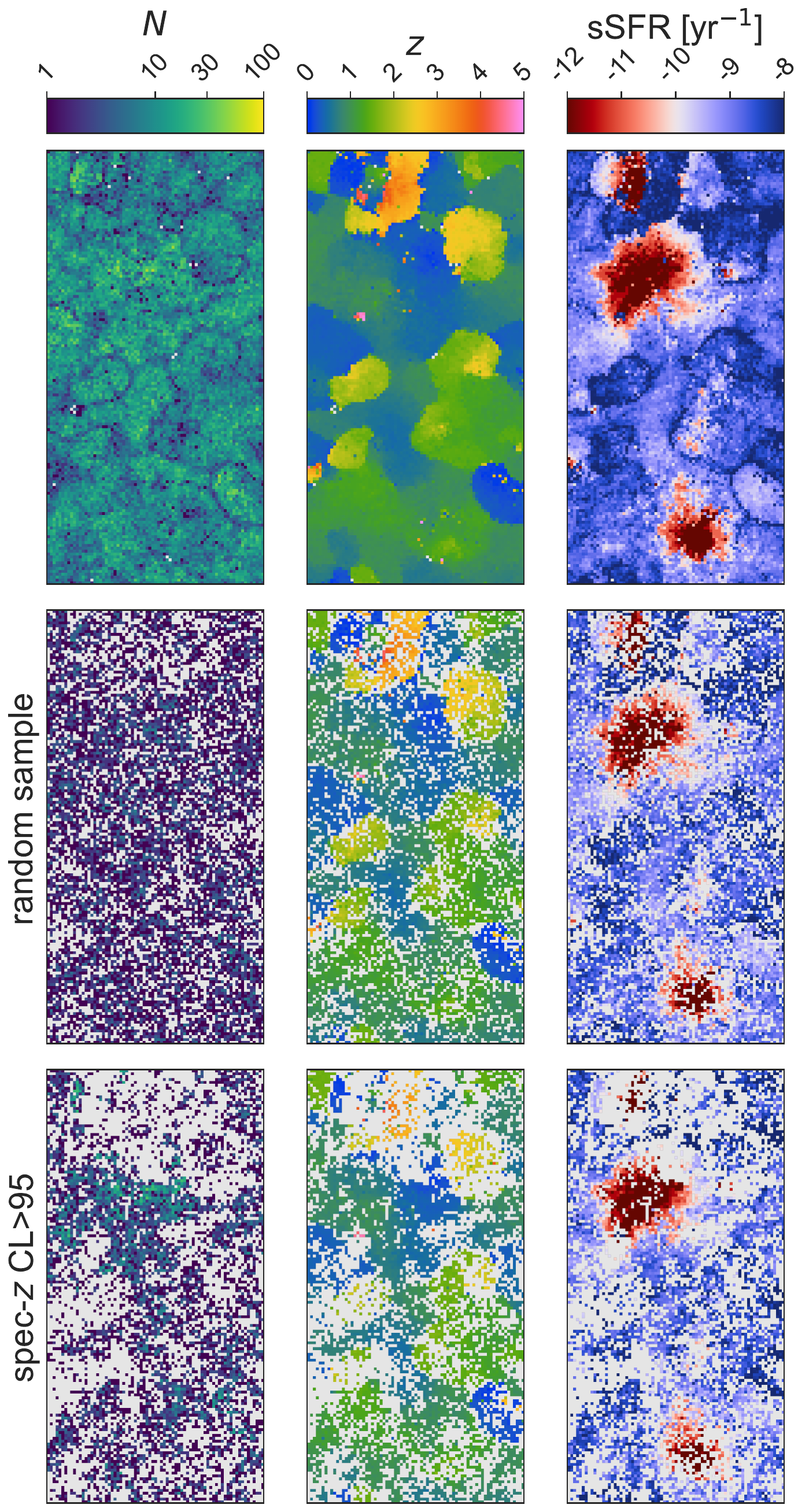}
    \caption{A self-organizing map (SOM) partitioning the $ugrizyJH$ color space, 
    color-coded by galaxy count (labeled $N$; left column), median redshift ($z$; center column), and median \lephare\ specific star formation rate (sSFR; right column). The top row shows the full sample (\added{112,387} sources) with the middle panel color-coded by \lepharez. 
    The middle row shows a random sample of photometric objects of the same size as the \specz\ sample (\added{14,477} sources) also color-coded by \lepharez\ \added{(center panel)}. The bottom row shows the \specz\ CL $>95$ sample (\added{14,477} sources), with the \specz\ coloring the middle panel. The middle row represents an ideally distributed training data case; it is representative of the full sample and gaps in the coverage of the color space are small and distributed more or less evenly. In contrast, the spectroscopic data (bottom row) illustrates the biased and incomplete coverage of color--redshift space that we aim to mitigate with UMAP-based techniques.
    }
    \label{fig:som_50_100}
\end{figure}

\figref{som_50_100} shows our SOM embedding of $ugrizyJH$ color space, color-coded by cell galaxy count, median redshift, and \added{median} sSFR in the left, center, and right columns, respectively. 
\added{This particular SOM embedding was generated with a random seed of 41, the same value used when optimizing hyperparameters. 
In the quantitative results described in \secref{redshift_estimation}, the mean performance metrics over 25 random seeds 0-24 are presented.} 
The top row shows the full photometric sample of \added{112,387} galaxies, while the middle and bottom rows are populated with \added{14,477} galaxies each. The middle row corresponds to a random subsample of the full dataset, while the bottom row corresponds to the subset of galaxies with high-confidence \speczs. Both redshift and sSFR are generally well-organized within particular neighborhoods of SOM cells, but there are often strong discontinuities in both quantities across adjacent cells at the borders of these neighborhoods. These discontinuities indicate that SOM offers a less reliable mapping between the observable colors and the physical parameters driving them, and introduces problems for tasks like redshift interpolation. Comparing the lower two rows illustrates the incomplete and  biased sampling of the color space by spectroscopic datasets. While the gaps in the random subsample are smaller and more dispersed, the spectroscopic sample lacks representation in large, correlated regions of the color space.

\subsection{UMAP} \label{sec:methods_umap}

While SOMs implement a discrete, (generally) rectangular binning of the data into a grid, UMAP is more flexible and projects each data point individually into a lower-dimensional space. The UMAP algorithm assumes the existence of a manifold on which the data is uniformly distributed and locally connected, and is optimized to preserve the topological structure of this manifold \citep{McInnes2018}.

UMAP constructs a weighted $k$-neighbor graph of the data that represents the data's connectivity, in which each point corresponds to a vertex and each vertex has edges to its $k$ nearest neighbors. 
In this graph, each point will have an edge weight of one for its nearest neighbor, indicating they are connected, with exponentially decreasing edge weights out to the $k$-th neighbor, indicating progressively lower probability of more distant points being connected. The edge with weight one ensures each point is connected to at least one other, while the decreasing edge weights to the $k$-th neighbor allow data with non-uniform density to be handled consistently.
UMAP determines the low-dimensional embedding by optimizing a weighted graph of the output space for similarity to that of the input space, with the difference that in this case the weights are a function of absolute distance in the low-dimensional space rather than the distance to the $k$-th nearest neighbor.

The critical hyperparameters when using UMAP are therefore the number of neighbors used to construct the graph (\nneighbors), 
the desired minimum distance between points in the output space (\mindist),
and the distance metric used in the input space (e.g., Euclidean, Manhattan, or cosine; the Euclidean metric is used in the output space). 

We implemented UMAP using the \texttt{umap-learn} Python package. As with SOM, we tested a variety of hyperparameter choices, including values of \nneighbors\ ranging from 10 to 200.
\added{As there is no reason to enforce a minimum distance (galaxies can be arbitrarily close in color space) and our aim is to work in a space densely sampled by the data, we set the \mindist\ parameter to 0.00.
We obtained the best results using the Manhattan distance metric in the input space and \nneighbors=70. These choices are explained in greater detail in \appref{umap_optimize}.}

\begin{figure*}
\centering
\begin{interactive}{animation}{figures/animation_260726.mp4.zip}
\includegraphics[width=\textwidth]{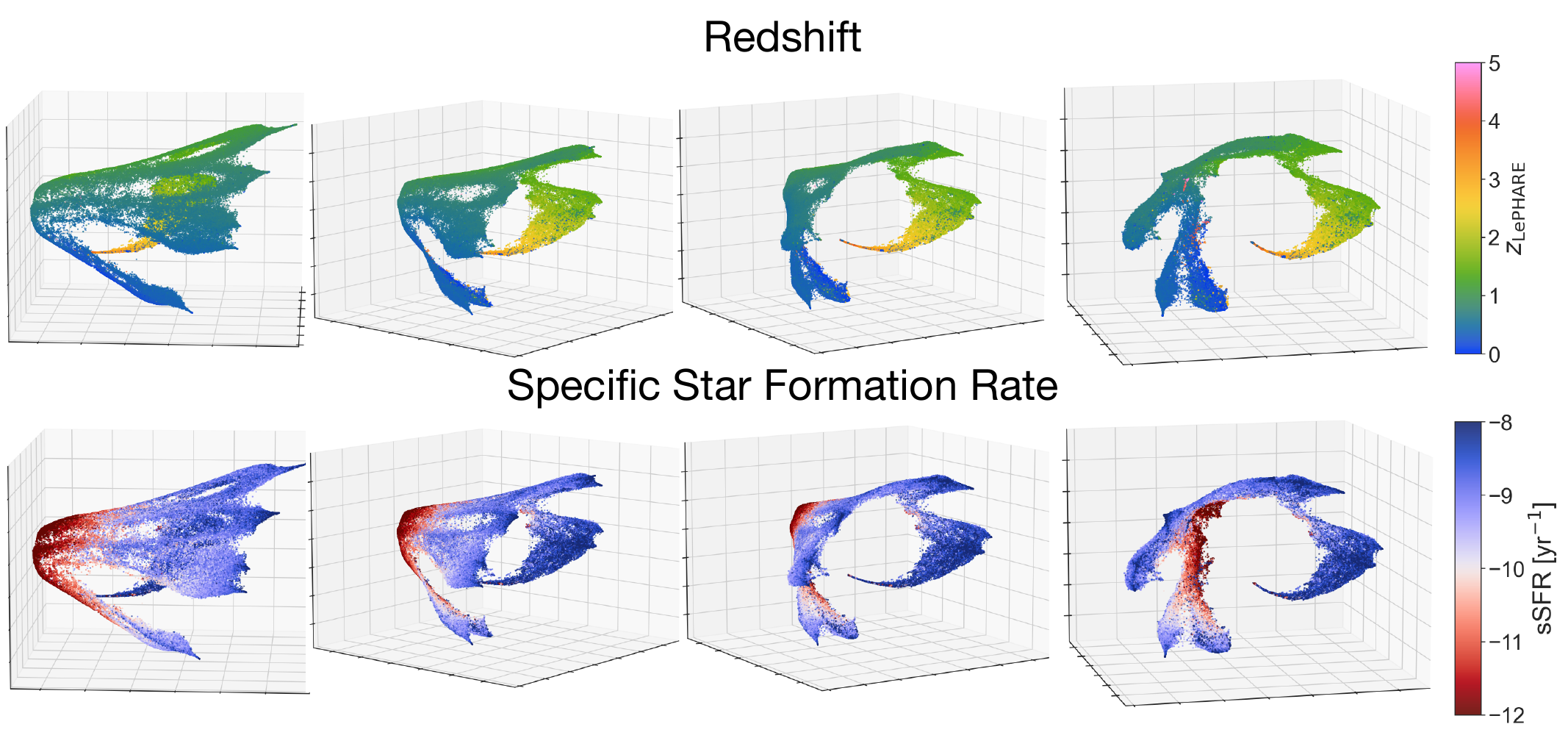}
\end{interactive}
\caption{A three-dimensional UMAP embedding of the $ugrizyJH$ color space for the full sample of \added{112,387} COSMOS2020 sources. The points are color-coded by \lephare\ \photoz\ (top) and sSFR (bottom). Objects lie on a nearly two-dimensional manifold in this space, exhibiting continuous, \added{monotonic, and locally orthogonal trends in redshift and sSFR.}
This figure shows selected frames from the thirteen-second animation available in the HTML version, in which the plots are rotating about the vertical axes to better illustrate the three-dimensional structure.}
\label{fig:umap_phot}
\end{figure*}

\begin{figure*}
\centering
\begin{interactive}{animation}{figures/animation_260726.mp4.zip}
\includegraphics[width=\textwidth]{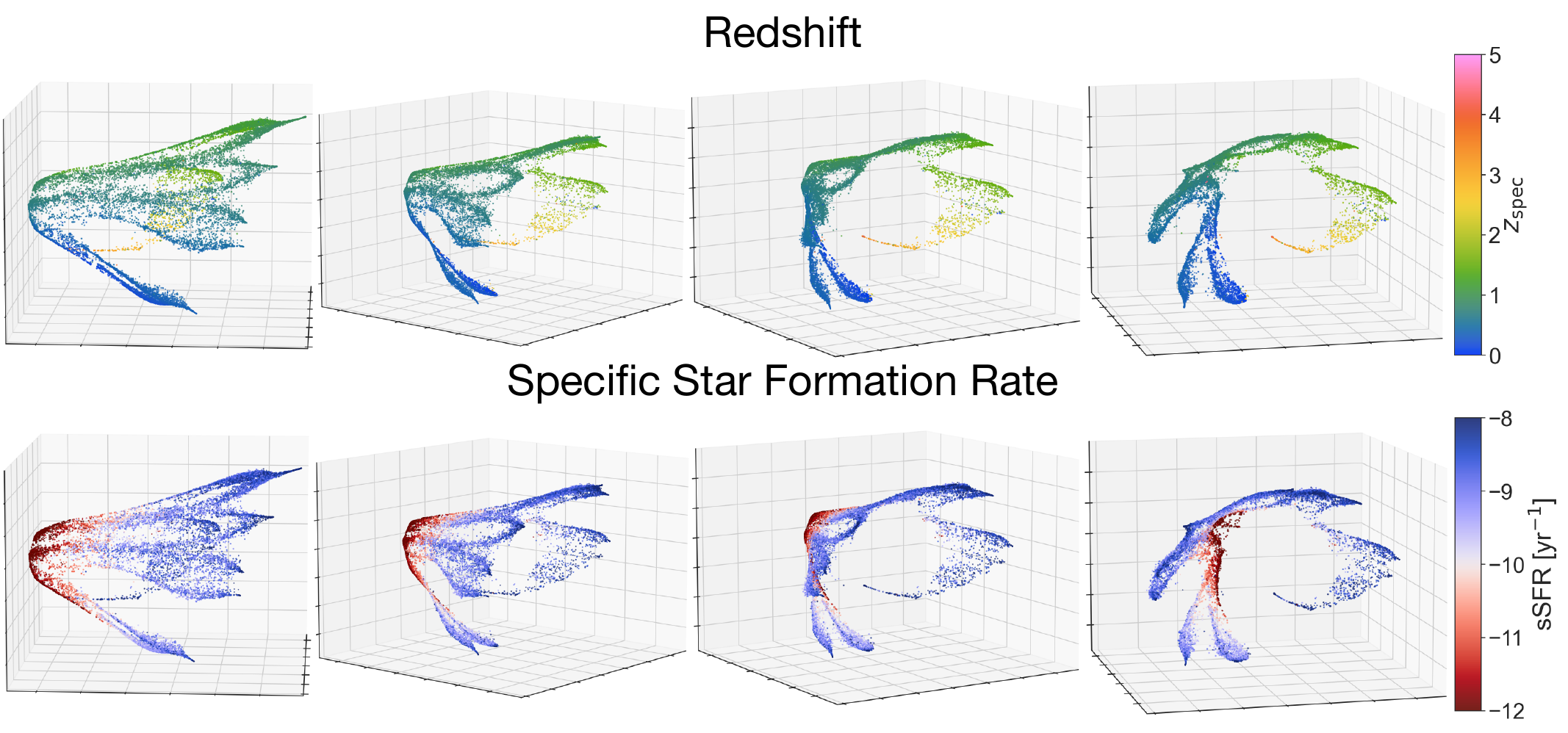}
\end{interactive}
\caption{A three-dimensional UMAP embedding of the $ugrizyJH$ color space showing only the \added{14,477} sources with high-confidence \speczs. The points are color-coded by \specz\ (top) and \lephare\ sSFR (bottom). 
Compared to the photometric objects (see \figref{umap_phot}), the spectroscopic objects provide a more sparse and biased sampling of the manifold.
However, unlike the SOM in \figref{som_50_100}, the trends in redshift and sSFR are continuous across the manifold, enabling accurate interpolation between the spectroscopic objects. This figure shows selected frames from the thirteen-second animation available in the HTML version, in which the plots are rotating about the vertical axes to better illustrate the three-dimensional structure.}
\label{fig:umap_spec}
\end{figure*}

In \figrefs{umap_phot}{umap_spec} we show our UMAP embedding of $ugrizyJH$ color space populated with the \added{112,387} galaxies in the photometric dataset and the \added{14,477} galaxies in the spectroscopic dataset, respectively, corresponding to the datasets in the top and bottom rows of \figref{som_50_100}. 
\added{This particular UMAP embedding was generated with a random seed of 41 and was the version used to select this set of hyperparameters. In the quantitative results described in \secref{redshift_estimation}, the mean performance metrics over 25 random seeds (0--24) are presented.}
We recovered a thin manifold embedded in three dimensions with continuous, monotonic, and roughly orthogonal variation in redshift and sSFR. Although the spectroscopic sources shown in \figref{umap_spec} provide a relatively sparse sampling of the space, the well-behaved and physically-meaningful organization of the manifold mitigates the impact of this. 

\added{By embedding each object into a continuous space, UMAP preserves more of the information contained in the detailed structure of the data in the input space than a discrete dimensionality reduction method can. As we demonstrate below, the smooth and monotonic mapping of UMAP coordinates to redshift allows one to estimate that quantity across the manifold from only a small and biased sampling of high-confidence \speczs. It is important to note, however, that the UMAP coordinates themselves are not expected to be physically meaningful and some level of information loss is inherent to dimensionality reduction. Preservation of global structure in the embedding can be highly dependent on the dataset used, the hyperparameter tuning, and the initialization of the output space \citep{Kobak2021, Marx2024}. We use UMAP's default ``spectral" initialization, which was found by \citet{Kobak2021} to improve global structure preservation compared to random initialization.}

\section{Redshift Estimation} \label{sec:redshift_estimation}
 
\subsection{Point Estimate Methods} \label{sec:interpolation}

Since our goal is to assess how well redshift can be predicted from position in the SOM or UMAP, we implemented simple algorithms for redshift estimation from each. We then use common \photoz\ evaluation metrics (see \secref{performance_metrics}) to quantify the reliability of the color--redshift mapping in these compressed spaces, allowing us to investigate the potential application of UMAP-based methods to optimizing training sets for \photoz\ algorithms. \added{To provide a comparison, we have also implemented a similar redshift estimator in the input $ugrizyJH$ color space used to generate the SOM and UMAP embeddings.}

For SOM, we first used the median of the training redshifts within a cell, hereafter referred to as \somz, as a redshift estimator. 
\added{However, this method can provide no redshift estimate for galaxies in cells with no training $z$'s. In these cases, we find the closest cell to a given galaxy that does have training redshifts based on minimizing the distance to the mean $ugrizyJH$ colors of the training objects in each cell. We label the combined algorithm which uses either the median of the training redshifts within the same cell if available or the median $z$ in the closest cell in the $ugrizyJH$ color space when needed as \nearestz. While both of these algorithms are SOM-based and use the same embeddings, we use \somz\ to refer specifically to estimates that exclusively use training redshifts within the same cell as a test object (and hence may provide no redshift prediction at all for a large fraction of the sample), and \nearestz\ for the SOM-based method with broader applicability. The \nearestz\ algorithm allows us to make a direct comparison to the other methods by obtaining a redshift estimate for every test object, while \somz\ allows us to separately evaluate redshift predictions based only on objects with training data in the same cell.}

For UMAP, we used the median redshift of the \added{15} nearest neighbors in the embedding, which we label \umapz\ (see \appref{umap_optimize} for more details). In each case we used the median rather than the mean due to its robustness to outliers (e.g., incorrect training redshifts); we also tested using the mean instead but obtained worse results,
\added{as described in \appref{optimize}.}
We note that the median $k$NN-based interpolation can easily be replaced with a more sophisticated method, as we intend to do in subsequent work. 
\added{For the $ugrizyJH$ color space we also implement a simple $k$NN-based estimator, taking the median redshift of the 11 nearest neighbors in the color space defined using the Manhattan distance metric (see \appref{colors_optimize} for more details).}

\added{
For all methods and both training scenarios, we quantify uncertainties due to the finite size of the test samples using bootstrap resampling. We also quantify uncertainties associated with hyperparameter optimization choices for each estimator by determining how performance metrics each change when a given hyperparameter is varied by $\pm15\%$. In the cases of the SOM and UMAP, we also assess the uncertainties associated with varying the random seeds used in generating the embeddings; for this purpose, however, we train with the full training samples and test with the test samples as no hyperparameter choices are involved. The procedures used are explained in more detail in \appref{uncs}. 
}

\subsection{Point Estimate Tests} \label{sec:point_estimate_tests}
\added{From our dataset of 112,387 sources we initially extracted a validation set of 14,477 galaxies, which is identical in size to the \specz\ sample but excluded all \specz\ sources. This validation set was used to optimize the hyperparameters of our embeddings and estimators as described in \appref{optimize}, as well as to quantify the uncertainties associated with this process as described in \appref{uncs}. We perform two different splits on the remaining 97,910 galaxies for our final quantitative testing. The first is the \lephare-trained case, for which the training set is a random subsample of equal size to the \specz\ sample with \lepharezs\ used as redshift labels, and the test set is the remaining 84,433 objects. The second is the \specz-trained case, for which we train with the 14,477 \specz\ objects and the test set consists of the remaining 83,433 objects. By reserving a validation set, we are able to optimize our embeddings and estimators without using the redshift labels of any objects involved in our final tests. The two separate training/test splits allow us to study the impact of using biased training data (the \speczs) as compared to an ideal situation where training data are fully representative (the \lephare-trained case).}




\added{When training with the \lepharezs, across the 25 SOM embeddings generated with different random seeds an average of about 25\% of the sample (21,030 sources) occupied cells that lacked training redshifts. When training with the CL $>95$ \speczs, this was the case for roughly 44\% of the sample (36,263 sources). These sources are neglected entirely in the \somz\ performance statistics, while in the case of \nearestz\ these sources are matched to their nearest neighbor in the input color space among the mean colors of each cell that does have training redshifts. 
}

\subsection{Performance Metrics} \label{sec:performance_metrics}
We assessed the redshift estimation performance by comparing the predicted redshifts to the \lephare\ \photozs\ of the test sample, either taken as a whole or when divided into redshift bins of width 0.2 in $z$.
In the following definitions, $\Delta z$ indicates the estimated redshift minus the ``true" (\lephare) redshift, $z$.
To evaluate our \photoz\ point estimate predictions, we used the fraction of outliers, normalized median absolute deviation, and prediction bias:

\begin{itemize}
\item \textbf{the fraction of outliers (\fout)}, defined as the fraction of predicted \photozs\ for which $| \frac{\Delta z}{1+z}|>0.15$;

\item \textbf{the normalized median absolute deviation (\nmad)}, defined as
\\ $1.4826\times$Median$(|\frac{\Delta z}{1+z}-$Median$(\frac{\Delta z}{1+z})|)$; and

\item \textbf{the prediction bias}, defined as $\langle \frac{\Delta z}{1+z} \rangle$.
\end{itemize}
In the case of the fraction of outliers, the threshold of 0.15 is commonly used in \photoz\ validation tests in the literature (e.g., \citealt{Hildebrandt2012}, \citealt{Weaver2022}, and \citealt{Khostovan2025}). We used \nmad\ as a robust metric for scatter, which is not affected by outliers so long as they are not too numerous.

\subsection{Global Performance}
\label{sec:global}
\begin{sidewaystable*}
\centering
\footnotesize

\begin{tabular}{c ccc ccc}
& \multicolumn{3}{c}{All Redshifts} &
\multicolumn{3}{c}{\zlp\ $>1.5$} \\
\cmidrule(r{0.5em}){2-4}
\cmidrule(l{0.5em}){5-7}

&
\fout\ & \nmad\ & Bias &
\fout\ & \nmad\ & Bias \\
\hline
\multicolumn{7}{c}{\lephare-trained} \\
\hline

\colorsz\ &
\textbf{0.0198 $\pm$ 0.0005} & \textbf{0.0182 $\pm$ 0.0002} & \textbf{0.0039 $\pm$ 0.0004} &
0.0377 $\pm$ 0.0017 & \textbf{0.0256 $\pm$ 0.0005} & -0.0251 $\pm$ 0.0010 \\

\somz\
\footnote{These statistics reflect cells with training data only.}
&
0.0194 $\pm$ 0.0026 (0.2666 $\pm$0.0035)
\footnote{\fout\ (all redshifts, \lephare-trained, values only for objects in cells used for \somz): 0.0122 $\pm$ 0.0004 (\colorsz) and 0.0134 $\pm$ 0.0006 (\umapz)}
& 0.0234 $\pm$ 0.0006 & 0.0049 $\pm$ 0.0013 &
0.0366 $\pm$ 0.0040 (0.3191 $\pm$0.0080)
\footnote{\fout\ (\zlp\ $>1.5$, \lephare-trained, values only for objects in cells used for \somz), 0.0230 $\pm$ 0.0014 (\colorsz) and 0.0233 $\pm$ 0.0016 (\umapz)}
& 0.0315 $\pm$ 0.0009 & -0.0187 $\pm$ 0.0018
\footnote{Bias (\zlp\ $>1.5$, \lephare-trained, values only for objects in cells used for \somz): -0.0170 $\pm$ 0.0010 (\colorsz) and -0.0155 $\pm$ 0.0012 (\umapz)}
\\

\nearestz\ &
0.0278 $\pm$ 0.0017 & 0.0247 $\pm$ 0.0005 & 0.0064 $\pm$ 0.0016 &
0.0520 $\pm$ 0.0032 & 0.0334 $\pm$ 0.0007 & -0.0268 $\pm$ 0.0021 \\

\umapz\ &
0.0218 $\pm$ 0.0005 & 0.0215 $\pm$ 0.0001 & 0.0064 $\pm$ 0.0005 &
\textbf{0.0377 $\pm$ 0.0016} & 0.0286 $\pm$ 0.0005 & \textbf{-0.0228 $\pm$ 0.0009} \\

\hline
\multicolumn{7}{c}{\specz-trained} \\
\hline

\colorsz\ &
0.0257 $\pm$ 0.0006 & \textbf{0.0223 $\pm$ 0.0002} & \textbf{0.0037 $\pm$ 0.0004} &
0.0509 $\pm$ 0.0013 & 0.0342 $\pm$ 0.0004 & --0.0362 $\pm$ 0.0008 \\

\somz\ 
\footnote{These statistics reflect cells with training data only.}
& 0.0215 $\pm$ 0.0026 (0.4468 $\pm$0.0034)
\footnote{\fout\ (all redshifts, \specz-trained, values only for objects in cells used for \somz): 0.0113 $\pm$ 0.0005 (\colorsz) and 0.0106 $\pm$ 0.0006 (\umapz)}
& 0.0236 $\pm$ 0.0006 & 0.0031 $\pm$ 0.0012 
\footnote{Bias (all redshifts \specz-trained, values only for objects in cells used for \somz): 0.0019 $\pm$ 0.0004 (\colorsz) and 0.0020 $\pm$ 0.0005 (\umapz)}
& 0.0727 $\pm$ 0.0071 (0.6787 $\pm$0.0079) & 0.0339 $\pm$ 0.0013 
& -0.0274 $\pm$ 0.0027
\footnote{Bias (\zlp\ $>1.5$, \specz-trained, values only for objects in cells used for \somz): -0.0235 $\pm$ 0.0016 (\colorsz) and -0.0208 $\pm$ 0.0019 (\umapz)}
\\

\nearestz\ &
0.0430 $\pm$ 0.0018 & 0.0276 $\pm$ 0.0004 & 0.0089 $\pm$ 0.0017 &
0.0972 $\pm$ 0.0043 & 0.0418 $\pm$ 0.0008 & -0.0396 $\pm$ 0.0021 \\

\umapz\ &
\textbf{0.0246 $\pm$ 0.0007} & 0.0250 $\pm$ 0.0001 & 0.0047 $\pm$ 0.0004 &
\textbf{0.0392 $\pm$ 0.0019} & \textbf{0.0334 $\pm$ 0.0005} & \textbf{-0.0309 $\pm$ 0.0012} \\

\hline
\end{tabular}
\caption{\label{tab:global} This table summarizes the performance of the various algorithms for the two training cases, across all redshifts (left) and for test objects with redshifts \zlp\ $>1.5$ only (right). The parenthetical values in the \somz\ \fout\ columns correspond to the values if all test objects in cells with no training data are considered outliers. The bolded values indicate the best algorithm based on the corresponding performance metric for that training case and redshift regime. In several cases, the lowest values of the performance metrics were obtained with \somz; however, we have bolded the next best values instead in these cases, as the \somz\ statistics ignore objects that were placed in cells without training data, and both \colorsz\ and \umapz\ perform better than \somz\ when restricted to the same set of objects. 
These cases and the corresponding results are detailed in the footnotes. The values for \nearestz\ ignoring objects in cells without training redshifts are by definition the same as the \somz\ values, so are not reported again.}

\end{sidewaystable*}


\added{
We present the overall performance statistics from each method and training scenario in \tabref{global}. In the cases of \somz, \nearestz, and \umapz, the presented values are the mean values of the performance metrics across 25 different realizations of each embedding, generated with random seeds from 0 to 24 and with hyperparameters fixed to the optimized values. The uncertainties on the performance metrics combine in quadrature the uncertainties associated with each hyperparameter choice, the finite test set size (as assessed by bootstrap resampling), and the variation of random seeds. For \colorsz, there is no embedding step, so the only uncertainties are due to the choice of $k$ value and the finite test set size. The calculation of the uncertainties is detailed in \appref{uncs}.
}

\added{
Unsurprisingly, \colorsz\ performs very well in the \lephare-trained scenario. There should inevitably be some information loss in the compression from seven dimensions to three, and the combination of training with a representative sample and a majority of the test dataset being in the densely-sampled $z<1.5$ regime make the limitations of working in a higher-dimensional space have less impact in this case.
The relatively good performance of \somz, particularly in the \specz-trained case, is also expected considering that for this algorithm the performance metrics simply ignore the objects in the regions of color space most affected by spectroscopic bias and incompleteness, which present challenges for the other algorithms in this test. As detailed in the footnotes of \tabref{global}, \colorsz\ and \umapz\ both yield superior performance metrics to \somz\ when the same objects are ignored, whereas \nearestz\ results are by construction identical to \somz\ in that case.
}

\added{
The relatively large gaps between the performance of \somz\ and \nearestz\ (particularly for \fout\ and bias) indicate that the redshift predictions for objects that share the same SOM cell are more reliable compared to those for objects that need to be predicted from other cells as their own lack training data. These objects generally belong to the regions of color space that are less densely sampled by the training set.
}

\added{
When restricting the statistics to only those objects with \zlp\ $>1.5$, the UMAP-based technique is particularly advantageous. In the \lephare-trained case, \umapz\ and \colorsz\ have nearly identical \fout\ performance, with \colorsz\ having somewhat better \nmad\ and \umapz\ having somewhat better bias. \umapz\ significantly outperformed \nearestz\ in this regime, and outperformed \somz\ as well when restricting the analysis to identical test samples. In the \specz-trained case at \zlp\ $>1.5$, the UMAP method has the greatest advantage over the other algorithms, and yields the best performance for all three metrics. For \fout, \umapz\ is the best by quite a large margin in this regime even in the lopsided comparison to \somz.
}

Overall, we find that using \umapz\ results in better
\added{redshift predictions}
in comparison to \somz\
\added{and \nearestz,}
with the difference in \fout\ being especially dramatic in the \specz-trained case
\added{and at $z>1.5$.}
\added{
While \colorsz\ performs best in the case with the most representative and densely sampled training data (\lephare-trained, all redshifts), \umapz\ does much better where the impact of spectroscopic bias and incompleteness is greatest (\specz-trained, \zlp\ $>1.5$).
}
Our results indicate that a UMAP-based approach to dimensionality reduction for mapping the color--redshift relation is particularly advantageous when using sparse or biased training data, as is the case with real-world spectroscopic datasets.

\subsection{Performance in Redshift Bins}

\begin{figure*}
\centering
\includegraphics[width=\textwidth]{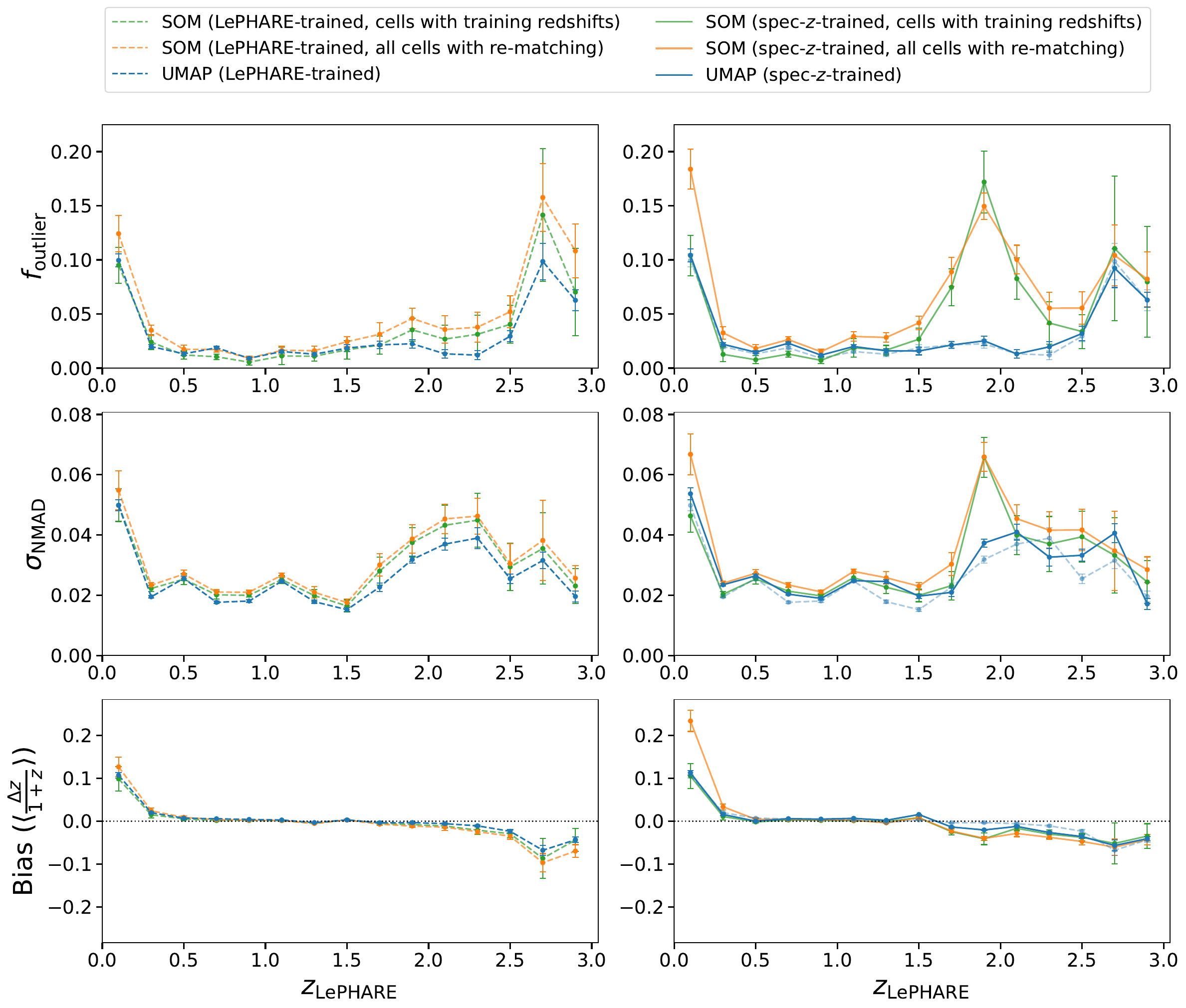}
\caption{\label{fig:binned_stat} \Photoz\ point estimate metrics \fout\ (top), \nmad\ (middle), and bias (bottom) calculated in $\Delta z=0.2$ redshift bins for
\added{\somz\ (green), \nearestz\ (orange), and \umapz\ (blue)}
when training with a random sample of COSMOS2020 \lephare\ \photozs\ (left) and the \specz\ CL $>95$ sample of \speczs\ (right).
\added{The \lephare-trained case for \umapz\ is shown on the right side as well at lower opacity for ease of comparison. The calculation of the error bars is described in detail in \appref{uncs}.}
\added{\umapz\ generally yields better performance metrics across the redshift range and especially at $z>1.5$. In the \specz-trained case, the difference is quite dramatic}
at $1.8<z<2.0$. Comparing the two UMAP curves in the right panels highlights the stability of the \umapz\ redshift prediction performance \added{when using two} very different training sets.}
\end{figure*}

In \figref{binned_stat}, we show the trends in \fout, \nmad, and bias with redshift for
\added{\somz, \nearestz, and \umapz.}
The panels on the left show the performance of the SOM- and UMAP-based estimators when training with the representative \lepharezs, while the panels on the right show the performance of the estimators when training with the realistically-biased \speczs. The exception is the dashed blue line representing the performance of the \lephare-trained UMAP method, which appears on both sides for ease of comparison. 

\begin{figure}
\centering
\includegraphics[width=\linewidth]{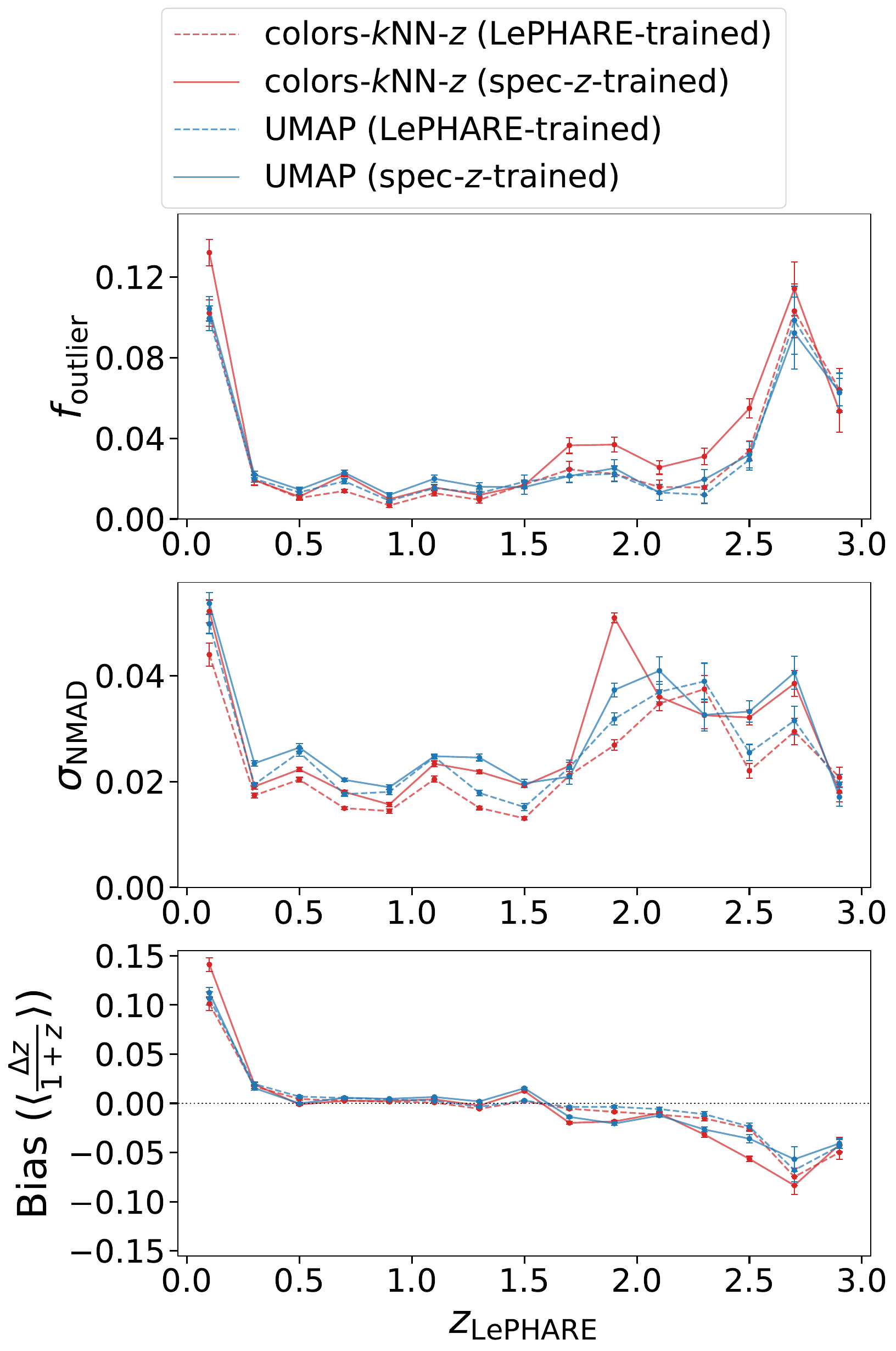}
\caption{\label{fig:binned_colors} 
\added{\Photoz\ point estimate metrics \fout\ (top), \nmad\ (middle), and bias (bottom) calculated in $\Delta z=0.2$ redshift bins for both \lephare-trained (dashed) and \specz-trained (solid) cases of \colorsz\ (red) and \umapz\ (blue). The calculation of the error bars is described in detail in \appref{uncs}. In the \fout\ plot, we see that while \colorsz\ performance degrades in the \specz-trained case in the lowest redshift bin and at $z>1.5$, the \specz-trained \umapz\ performance remains comparable to the \lephare-trained results. The \specz-trained case of \colorsz\ shows a spike in \nmad\ in the $1.8<z<2.0$ bin and an increased bias in both the lowest redshift bin and the $2.0<z<3.0$ regime. In comparison to \colorsz, the \umapz\ performance is highly resilient to the sparse and biased \specz\ training data.}
}
\end{figure}

\added{
In the \lephare-trained case, the SOM- and UMAP-based approaches perform similarly in both \fout\ and bias in the $0.5<z<1.5$ regime, while \umapz\ does somewhat better in the lowest redshift regime and significantly better in the $z>1.5$ regime (especially in \fout). \umapz\ yields better \nmad\ values across the entire redshift range, in addition to generally having smaller uncertainties on all metrics, primarily due to having less sensitivity to random seeds and hyperparameters (see \appref{uncs} for details). 
}

In the \specz-trained case, the impact of the spectroscopic data's biased sampling of the color space is clear in the performance of the SOM-based \added{techniques}, which is degraded considerably. 
\added{In contrast, the performance of \umapz\ is essentially unchanged in comparison to the \lephare-trained case, as seen in the similarities in the two blue lines. In the lowest-redshift bin, the advantage of \umapz\ over \nearestz\ is much more significant in all metrics. Most strikingly, both SOM-based techniques suffer a notable degradation in redshift prediction accuracy in the $1.5<z<2.5$ regime, especially in the $1.8<z<2.0$ bin where the \nearestz\ \fout\ is roughly an order of magnitude larger than that for \umapz. Additionally, as in the \lephare-trained case the uncertainties on performance metrics are generally smaller for \umapz\ than either SOM-based technique, as detailed in \appref{uncs}.
}

\added{In contrast to the SOM-based methods, the UMAP-based predictions} are minimally affected by the \added{spectroscopic} data's limited sampling of the color space, as illustrated in the similarity between the solid and dashed blue curves in the panels in the right column of \figref{binned_stat}. Furthermore, the spike in \fout, \nmad, and bias for the SOM-based approach at $z_\mathrm{LePHARE}\sim1.9$ is completely absent for the UMAP-based approach. This is the key benefit of UMAP's ability to capture the continuous structure of the photometric data driven by underlying physical parameters, such that the redshift of the objects varies smoothly and coherently across the low-dimensional UMAP embedding and can therefore by interpolated reliably.

\added{In \figref{binned_colors}, we compare the trends in the performance metrics as a function of redshift from \colorsz\ and \umapz\ for both training scenarios. While \colorsz\ does well in the \specz-trained case in the regimes least impacted by bias and incompleteness (i.e., $0.5<z<1.5$), the \nmad\ spikes at $1.8<z<2.0$ and the bias is worse for \colorsz\ in the lowest-$z$ bin as well as several of the highest-$z$ bins. In contrast, as was seen in \figref{binned_stat}, the performance of \umapz\ is minimally affected by the difference between the representative \lephare\ training data and the biased \specz\ training data. 
}

Overall, we find the UMAP-based approach to redshift estimation to be highly resilient to biased spectroscopic training data, in contrast to \added{\colorsz\ and particularly the SOM-based approaches}. 
\added{\figref{binned_stat} shows large and statistically significant differences in the performance of the SOM-based methods between the two training cases, especially at \zlp\ $>1.5$, while these differences are much smaller for \umapz\ (although statistically significant for some bins given the UMAP method's much smaller uncertainties; see \figref{compare_uncs}). Similarly, Table \ref{tab:global} and Figure \ref{fig:binned_colors} show that \colorsz\ also suffers greater performance degradation when training with \speczs\ compared to \umapz, with the differences again being statistically significant.
} 

As visualized in \figref{som_50_100}, the discrete nature of the SOM and the discontinuity of the mapping to redshift across SOM cells make redshift estimation \added{difficult} in large regions of the color space when using non-representative training data. With UMAP, however, as visualized in \figref{umap_spec}, the continuity of the space and the smooth variation in redshift mitigate the issue of the incomplete sampling of the color space by spectroscopic data. \added{As shown in \figref{binned_colors}, \umapz\ delivers performance that is superior to \colorsz\ in the \specz-trained case at the redshifts which are most impacted by spectroscopic bias and incompleteness.  This further illustrates UMAP's ability to capture physically-meaningful information contained in the distribution of the data in the input color space while compressing to a lower-dimensional representation in which training sets will be more dense.}

\section{Discussion} \label{sec:discussion}

In this work, we found that using UMAP to compress the space of observed galaxy colors has several significant advantages over using a SOM. \added{Applying common photo-z evaluation metrics, we find that redshift can be predicted more accurately from UMAP coordinates than from a SOM, with the differences being highly statistically significant.}
Due to the discrete nature of the SOM, there is a tradeoff between the resolution with which the color space is represented and the numbers of sources in the cells, with higher cell counts representing the color space in greater detail but with increasingly sparsely populated (or even empty) cells. By projecting each source individually into the output space, UMAP avoids this tradeoff and preserves more information about the structure of the input space. 

SOM's enforced geometry and drive to assign sources to all cells inhibit it from recovering structure occupying a subset of the output space. By contrast, UMAP is able to recover a thin, almost two-dimensional manifold embedded in three-dimensional space. Similarly, UMAP is able to separate low- and high-redshift regions with empty space, while in the SOM projection they can occupy adjacent cells. 

Critically, the UMAP embedding exhibits continuous and monotonic trends in redshift and sSFR. This is extremely useful for interpolation across the manifold for redshift prediction, especially for regions with sparse training data. 
Our results indicate that UMAP dimensionality reduction can recover the simple, intrinsic structure of observed galaxy color space in a way that maps meaningfully to the underlying physical parameters driving it. 

The structure of the low-dimensional manifold recovered based on the full photometric sample maps cleanly enough to redshift that \umapz\ can reliably estimate redshifts based only on a small, highly-biased subset of the data. Furthermore, although some detailed structure contained in the higher-dimensional input color space has necessarily been lost or distorted in the compression process, $k$NN-based redshift estimation in the UMAP space yields a lower \fout\ than that in the seven-dimensional color space in the \specz-trained case. In the $z>1.5$ regime, which is sparsely sampled by the \specz\ training data, UMAP yields dramatically better redshift predictions, with \added{23\% lower \fout\ (0.0392 versus 0.0509), 2.3\% lower \nmad\ (0.0334 versus 0.0342), and 15\% lower bias magnitude (-0.0309 versus -0.0362).}
This is further indication that a UMAP-based approach to mapping color to redshift is particularly advantageous in the case of sparse and highly-biased training data.

Interestingly, \citet{Dey2022} recovered UMAP embeddings with key similarities to \added{those} presented in this paper using a very different approach. That investigation used a deep capsule network to compress SDSS $64\times64$-pixel galaxy postage stamp images in $ugriz$ bands ($64\times64\times5$ input) into 16-dimensional vectors, which were then further compressed into two- or three-dimensional UMAP embeddings. Similarly to our results, the authors in all cases obtained a curved manifold with a continuous and monotonic redshift sequence. Using Galaxy Zoo spiral/elliptical classifications available for these galaxies, they found morphological trends \added{locally} orthogonal to the trend in redshift. 
Given the correlation between galaxy morphology and sSFR, this is somewhat analogous to the trend in sSFR being \added{locally} orthogonal to the redshift sequence here. The similarity of our results further indicates that observed high-dimensional photometric galaxy data is largely structured by a small number of physical parameters, and this structure can be studied and leveraged using dimensionality reduction and machine learning.

Going forward, there are several applications of the UMAP-based technique we are exploring. Replacing our initial $k$NN-based interpolation with more sophisticated techniques (robust Gaussian process regression) will enable improved redshift estimation and estimation of redshift probability distribution functions and errors at any point in the manifold. Incorporating this, we will be able to interpolate across a UMAP-generated embedding of a photometric sample of galaxy color space populated with spectroscopic sources, and sample from this interpolated manifold to generate representative training data for \photoz\ algorithms. Likewise, this can be used to identify spectroscopically under-observed regions of color space for prioritization of future observations (e.g. \citealt{Masters2015}, \citealt{Khostovan2025}). The embedding can also be populated with mock/simulated galaxies to diagnose differences between their distribution and that of real data, and can similarly be used to improve template-based techniques. Finally, this approach can easily be extended to investigating galaxy properties other than redshift, such as mass-to-light ratio, stellar mass, and sSFR. We provide the necessary Python code to reproduce our analysis including all figures and animations on Zenodo under a Creative Commons Zero v1.0 Universal license: \dataset[doi:10.5281/zenodo.21716802]{https://doi.org/10.5281/zenodo.21716802}.

\begin{acknowledgments}
F.~A., J.~A.~N., and B.~H.~A.\  acknowledge the support of the National Aeronautics and Space Administration under grant No.~80NSSC24K0083.

B.D.\ is a postdoctoral fellow at the University of Toronto in the Eric and Wendy Schmidt AI in Science Postdoctoral Fellowship Program, a program of Schmidt Sciences.

The presentation of this work was influenced by valuable conversations with colleagues at the University of Pittsburgh and other institutions, including Joel Leja, Joshua Speagle, Tianqing Zhang, Ashod Khederlarian, Yoquelbin Salcedo Hernandez, and Federico Berlfein. We would also like to thank the Survey Science group at the Kavli Institute for Cosmological Physics at the University of Chicago and its co-leader Chihway Chang, who hosted the author to present early results during the writing of this paper and offered valuable feedback and lines of questioning. \added{We also thank the anonymous referee and editors for helpful suggestions which have improved this paper.}



\end{acknowledgments}

\software{
\texttt{astropy} \citep{astropy},
\texttt{matplotlib} \citep{matplotlib}, 
\texttt{minisom} \citep{minisom}, 
\texttt{numpy} \citep{numpy}, 
\texttt{pandas} (\citealt{McKinney2010}; \citealt{pandas}), 
\texttt{scikit-learn} \citep{scikit-learn}, \texttt{scipy} \citep{scipy}, 
\texttt{seaborn} \citep{seaborn},
\texttt{umap-learn} \citep{McInnes2018}
}

\appendix

\section{\added{Optimization of Algorithms}} \label{app:optimize}

\added{In this appendix, we provide details on how we optimized the hyperparameter choices for each algorithm used in the paper.  In all cases, quantitative tests of the optimizations were performed using the validation set described in Section \ref{sec:methods_som}}.

\subsection{\added{\colorsz}} \label{app:colors_optimize}

\added{The \colorsz algorithm requires the fewest choices since no dimensionality reduction step is involved. Optimizing this method amounts to choosing a distance metric to use in the color space when defining neighbors, choosing a value for the number of neighbors to use ($k$) itself, and deciding whether to use the mean or median of the redshifts of the $k$ nearest neighbors. As in the optimization of the other algorithms, only the validation set was used in the testing that guided decisions on parameters.}

\added{In this and the following sections of this appendix, all calculations for the validation set were performed using a variant of five-fold cross-validation.  Specifically, we randomly divided the validation sample into five subsets.  For each subset, we predicted redshifts using a given method when training with the remaining 80\% of objects, and then calculated our standard test statistics (\nmad, \fout, and bias). We then averaged the values from all five subsets to estimate each parameter, and used the spread amongst the results from each subset to assess the uncertainties in these values. This method allowed metrics to be calculated using the entire validation set (reducing the uncertainties in them) while avoiding overfitting by ever training and testing with the same data.}

\added{Using the validation set, we tested the results from predicting redshifts from the mean or median $z$ of the $k$ nearest neighbors defined using either the Manhattan or Euclidean distance metric, with each statistic computed using odd values of $k$ ranging from 3 to 27. For both distance metrics, using the median gave much better results for \fout\ and \nmad\ than using the mean, with somewhat better bias. When employing the median, the Manhattan metric yielded significantly better \fout\ for most $k$ values, somewhat better \nmad, and similar bias to the Euclidean metric. The \fout\ and bias values were weakly dependent on $k$ and lacked any clear trend, while the \nmad\ was strongly dependent on $k$, exhibiting a sharp drop with increasing $k$ up to a clear minimum at $k=11$, with a weak increase with $k$ at higher values. Therefore, we optimized our \colorsz\ implementation by using the median of the redshifts of the $k=11$ nearest neighbors, as identified using the Manhattan distance metric.}

\subsection{\added{\somz}} \label{app:som_optimize}

\added{We restricted our SOM embeddings to consist of rectangular cells with a $1\times2$ aspect ratio, as used in, e.g., \citet{Masters2015}. Specifically, we tested maps of dimensions (8, 16); (12, 24); (18, 36); (25, 50); (35, 70); (50, 100); and (75, 150). These size intervals were chosen such that the total cell number roughly doubles with each increase. Each SOM cell corresponds to a vector in the input $ugrizyJH$ color space, with data grouped to the nearest vector and assigned to the corresponding cell in the grid at the end of training. The learning rate hyperparameter initializes how far the vectors move during each training step, while the sigma hyperparameter initializes the spatial extent to which neighboring vectors are pulled along as one updates. The values of both hyperparameters decay as the embedding converges, with larger input values enabling more extensive re-ordering of the vector positions early on and therefore more emphasis on representing the global structure. 
For the learning rate, we tested values from 0.2 to 1.0 in intervals of 0.2. 
For sigma, we tested values of 0.5, 1.0, 2.0, and 4.0. We generated SOM embeddings with these hyperparameter combinations using either the Euclidean or Manhattan distance metric in the input color space. 
In all cases, we used two million training epochs.}

\added{The embeddings were in all cases determined using the full photometric dataset. We then tested the ability to predict redshift from each embedding using five-fold cross-validation with our validation dataset. We also inspected the SOM embeddings visually, color-coding by either the density within each cell (from the full sample) or by the median cell redshift (for the validation sample only). As for the optimization of other algorithms, only the redshift labels of validation set objects were used for both qualitative and quantitative testing. Either the mean or median redshift of the training objects in a cell could be used as an estimator; in the case of \nearestz\ the difference in performance between the two was much less significant than in the cases of \colorsz\ and \umapz. For \fout\ and \nmad\ the median yielded smaller values, while the bias results were very similar. As such, we used the median cell training redshift in the optimization described below and in the results reported throughout the paper.}

\added{Although we report performance statistics for redshift prediction from SOMs either when restricting to cells with training data or for all test objects after re-matching, we did not optimize separate embeddings for the two cases; rather, we only used \nearestz\ (i.e., prediction with re-matching) when optimizing, as this algorithm can most cleanly be compared to \colorsz and \umapz.  Simply ignoring the objects in cells without training redshifts, as the \somz\ algorithm does, runs counter to our goal of addressing the problems of spectroscopic bias and incompleteness in \photoz\ training sets, as it amounts to dropping the most affected regions of color space. This makes \nearestz\ results of primary interest for us.  However, by using the same SOM mapping for \somz\ and \nearestz\, we are able to compare the two algorithms to each other most directly.}



\added{However, we separately optimized embeddings for the Euclidean and Manhattan distance metrics. Generally, higher sigma and learning rate values led to faster convergence and to embeddings with larger regions featuring relatively smooth gradients in redshift across cells. Similarly, the embeddings with lower total cell counts also converged more quickly. One might expect that the smoother redshift gradients from embeddings with higher sigma and learning rate would generally lead to better redshift predictions, but quantitative testing revealed that this was not the case. We speculate that this may be due to a tradeoff between capturing global and local structure in the mapping of color space to redshift, with cases of high sigma and learning rate leading to an excessive focus on global trends. We chose a learning rate of 0.8 as the best value for both distance metrics, which led to well-optimized, converged embeddings (i.e., with a well-distributed density of objects across cells, neighboring cells generally having similar median redshifts, and good performance in the cross-validation testing) as well as reasonable run times.}

\added{Generally, the performance of redshift interpolation in these SOM embeddings was not very sensitive to changes in the hyperparameters. The exceptions occurred for the two smallest SOM sizes, which delivered drastically worse redshift prediction compared to the others, and the most extreme values we tested for sigma and learning rate. Excessively high values of these hyperparameters worsened performance, while excessively low values yielded poorly-converged embeddings, with many or even most cells containing no objects and with patches of a few cells at most surrounded by empty ones.}

\added{When the Euclidean distance metric was used, the embedding with the best redshift prediction performance was of size $50\times100$ cells with sigma and learning rate values of 1.0 and 0.8, respectively. Comparing this quantitatively to the Manhattan embedding of the same size and hyperparameters, the Manhattan case had slightly better mean performance metrics with considerably smaller scatter between folds. For the Manhattan metric, the best embedding was of size $75\times150$, again with sigma of 1.0 and learning rate of 0.8, and this was our best embedding overall. Based on these results, our \somz\ and \nearestz\ implementations use $75\times150$ grids, the Manhattan distance metric, sigma of 1.0, learning rate of 0.8, and the median redshift of training objects in a cell as a redshift estimator.}

\subsection{\added{\umapz}} \label{app:umap_optimize}

\added{We generated UMAP embeddings with either two or three dimensions and a range of hyperparameters 
using the \texttt{umap-learn} Python package. Other than the dimensionality of the UMAP embedding and the key hyperparameters \mindist\ and \nneighbors, options included whether to rescale the input data and whether to use the densMAP version of UMAP \citep{Narayan2020}. While UMAP normally assumes the existence of a manifold on which the data is uniformly distributed, the densMAP version aims to preserve variations in local density of the input data; that algorithm was used in \citet{Dey2022}.}

\added{We used the \texttt{scipy StandardScaler} function to rescale our color vectors to have means of zero and unit variance, and performed median $k$NN-based redshift prediction in the rescaled color space itself as well as in UMAP or densMAP embeddings generated from it (with \nneighbors=80, \mindist=0.00, and a Manhattan metric) for a range of $k$ values. In color space the Manhattan metric was used to identify neighbors, while in the UMAP and densMAP spaces the Euclidean metric was used for that purpose (although a Manhattan metric was used when generating the embeddings). Redshift prediction from the UMAP space generated from non-rescaled data vastly outperformed that which resulted from using the standard-scaled data in both \fout\ and \nmad, with somewhat better results for bias as well (significantly so for higher $k$ values). The non-rescaled data similarly yielded significantly better results in the densMAP embeddings, for all metrics and at all $k$ values tested.  Comparing the UMAP and densMAP embeddings to each other, the UMAP embedding with non-rescaled data yielded the lowest \fout\ of the four by a large margin across all $k$ values. This embedding also yielded significantly smaller \nmad\ than the others, a significantly lower bias than the UMAP and densMAP spaces generated from rescaled colors, and marginally better bias than the densMAP generated from non-rescaled colors. For \fout, the UMAP embeddings both performed better than the densMAP embeddings, with the non-rescaled version yielding the lowest values for each embedding approach. For \nmad\ and bias, the gap between the rescaled and non-rescaled versions was larger than the gap between UMAP and densMAP, with the non-rescaled version being better for both algorithms at all $k$ tested.}

\added{Direct redshift prediction from coordinates in the non-rescaled $ugrizyJH$ color space likewise outperformed the rescaled version. The non-rescaled space yielded significantly lower \fout\ for most $k$ values, and at least somewhat better \fout\ for all $k$ tested. The \nmad\ and bias results were similar between the two color spaces. We speculate that the superior performance with non-rescaled data indicates that the different dynamic ranges of the different colors contains information useful to redshift prediction. }

\added{Of course, there exist other schemes for rescaling the input color data before performing dimensionality reduction or other machine learning tasks including redshift predictions, as in, e.g., Appendix A of \citet{Yin2025}. 
An exhaustive investigation of  possible rescaling methods is beyond the scope of this work.
Based on the results of the tests described here, we did not rescale the input color data when performing further optimization or in the final \umapz\ algorithm. }

\added{We next downselected amongst UMAP algorithm variations by visually comparing the resulting manifolds. Specifically, we assessed how smooth any trends in redshift were and whether the low- and high-redshift ends of the manifold were separated in the dimensionality-reduced space. As for the optimization of the SOM embeddings, we only used the redshift labels of validation set objects in these visualizations (although, as with SOM, the underlying dimensionality reduction was constructed using all objects with color measurements).}

\added{Using these criteria, the best results were obtained using three-dimensional UMAP embeddings generated from non-rescaled data, without using densMAP, using the Manhattan distance metric in the input space, and setting the \mindist\ hyperparameter to 0.00. Although we expected \mindist=0.00 to be the best choice for our approach since galaxies can be arbitrarily close in color space, we did test values up to 0.20; however, 0.00 was the only one that yielded embeddings without filamentary structure connecting the low-$z$ and high-$z$ ends of the manifold. We also tested values of \nneighbors\ from 10 to 200, and found that the structure of embeddings remained very similar for values between 50 and 90. }

\added{At this stage we initiated quantitative testing using the validation dataset, similarly to the methods used to optimize \colorsz\ and SOM hyperparameters. We first optimized the $k$ value to use for redshift prediction, adopting a fixed UMAP embedding generated with \nneighbors=70, \mindist=0.00, the Manhattan metric, and a random seed of 41. The Manhattan metric and \mindist=0.0 were adopted based upon our visual inspections, while \nneighbors=70 was at the center of the range over which results were nearly indistinguishable. The random seed of 41 was chosen as this value was not used in our later testing or quantification of the uncertainties associated with different random seeds. We used either the median or mean of the training redshifts of the $k$ nearest neighbors as our estimators, finding that the median yielded vastly superior results for \fout\ and \nmad, and somewhat better results for bias. Much as for \colorsz, there were no strong trends in \fout\ or bias with $k$, while \nmad\ declined steeply with increasing $k$ for small values, reaching a minimum at $k=15$ and then rising again at large $k$. }

\added{Having fixed $k=15$ on this basis, we measured redshift prediction errors while varying \nneighbors\ in increments of five from 50 to 90. In general, we found the performance to be only very weakly dependent on \nneighbors.
For \fout\ and bias, the variations between the values for different cross-validation folds were much larger than differences as a function of \nneighbors. 
For \nmad, there was a weak but detectable trend of worsening performance at both extremes of the range of \nneighbors, leading us to adopt 70 as our optimal value.
Given all these results, we use the Manhattan distance metric, \mindist\ of 0.00, and \nneighbors\ of 70 to generate our UMAP embeddings, with the median of the redshifts of the $k=15$ nearest neighbors in the embedding used to predict $z$.}

\section{\added{Uncertainty Quantification}} \label{app:uncs}

\added{
For all methods and both training scenarios, we quantify uncertainties due to the finite size of the test samples using bootstrap resampling. For the SOM and UMAP methods, we also assess the uncertainties associated with varying the random seeds used in generating the embeddings. We use the full training and test samples for each training case to estimate these uncertainties as no optimization choices are involved. In each case, the standard deviations of the performance metrics after 25 runs are taken as the uncertainties attributable to these sources.
}

\added{
We optimized each algorithm's hyperparameters quantitatively using the values of the performance metrics described in Section \ref{sec:performance_metrics} for the validation set. This included hyperparameters that affect the generation of the SOM and UMAP embeddings (SOM grid size, learning rate, and sigma; UMAP \nneighbors) as well as the $k$ values used for \umapz\ and \colorsz. We performed a variant of five-fold cross-validation, estimating the uncertainties in performance as a function of hyperparameter value based on the spread among the performance metrics across the five subsets. 
}

\added{In this optimization testing, there was generally not a single hyperparameter value that simultaneously optimized the three performance metrics, and often there were no clear minima at all (i.e., no clear trend and changes in performance from one hyperparameter value to the next having overlapping error bars). For the cases where there were identifiable minima (e.g., in \nmad\ for different $k$ values in \colorsz\ and \umapz), the width corresponded to roughly 15\% of the hyperparameter value. In order to estimate how much a given performance metric value might change given a different choice for which metric to prioritize in  optimization, we evaluated how much it changes under a $\pm15\%$ variation in the value of a given hyperparameter, taking the half-range of the means (over the five folds) as the uncertainty on that performance metric associated with that hyperparameter.
This allowed us to account for the ambiguity in the hyperparameter optimization in a consistent manner across the redshift estimators and quantify their sensitivity to these variations. The details of this calculation and the particular hyperparameter values used are provided in this appendix.
}

\subsection{\added{Uncertainties in Statistics for \colorsz}} \label{app:colorsz_unc}

\begin{figure*}
\centering
\includegraphics[width=\textwidth]{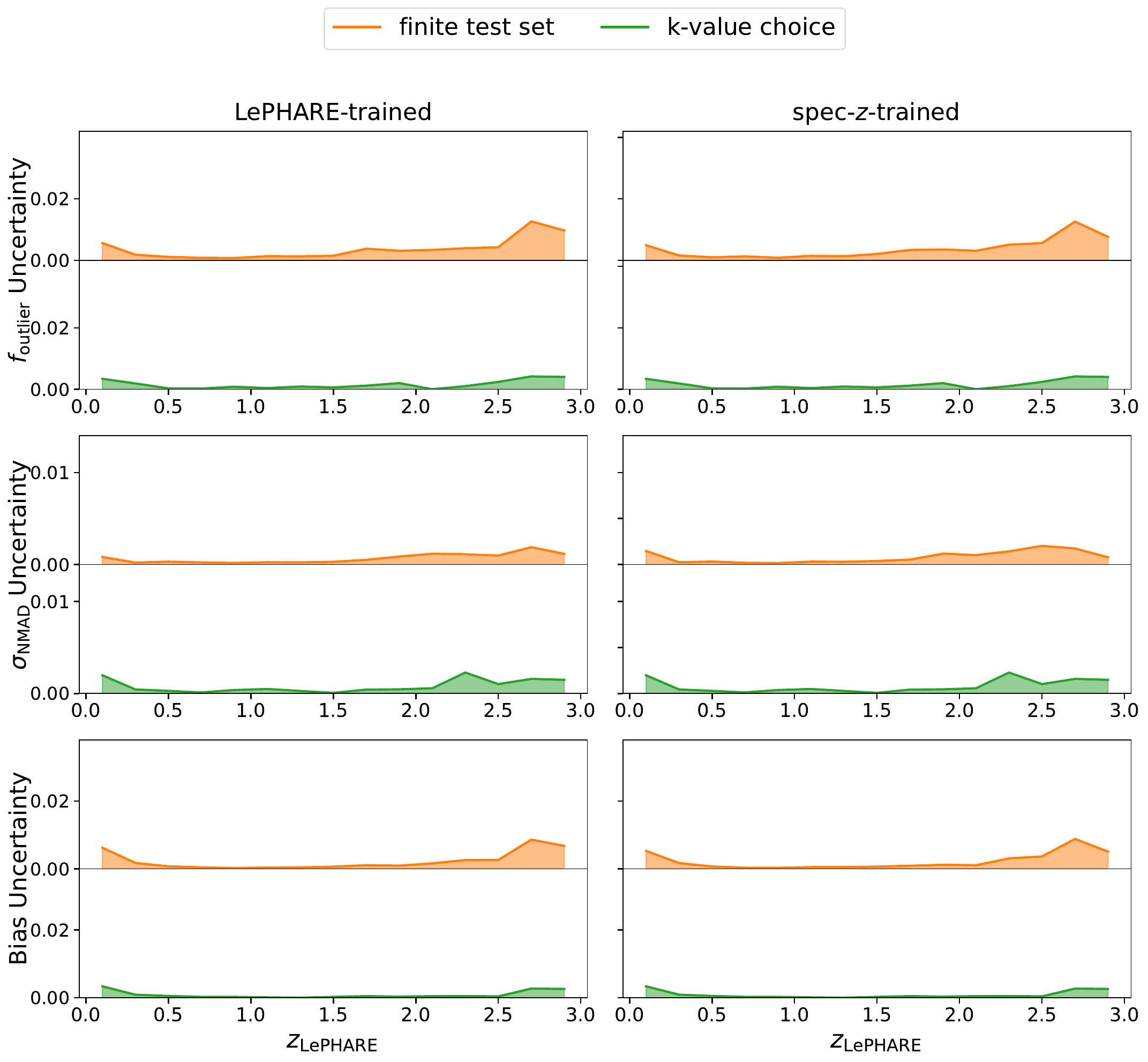}
\caption{\label{fig:uncs_colorsz}
\added{Contributions to the uncertainties in the performance metrics of \colorsz\ predictions  (i.e., those predicted from the median redshift of the 11 nearest neighbors in color space; see \secref{interpolation}) as a function of redshift. Since \colorsz\ does not involve a dimensionality reduction step like the SOM- and UMAP-based approaches, there are no uncertainties associated with a random seed or hyperparameters used to generate an embedding. The sizes of the \colorsz\ error bars in \figref{binned_colors} correspond to the quadrature sum of the uncertainties plotted here. 
Consistent colors are used for uncertainty sources shared across any of the four redshift predictors; e.g., orange is employed for the uncertainty associated with the finite test set, which applies to all methods.}
}
\end{figure*}

\added{As the \colorsz\ predictions (described in \secref{interpolation}) do not involve any dimensionality reduction step, they have the fewest potential sources of uncertainty of the methods used in this paper. To calculate the uncertainty associated with the choice of $k$ value, we used five-fold cross-validation with the validation dataset, following the procedure described in \S \ref{app:colors_optimize}.
For each numerical hyperparameter used in this and the following sections, we have quantified the uncertainties in performance statistics associated with the choice of its value by varying it by $\pm15\%$ and determining how a given statistic changes. For \colorsz\ the optimized $k$ value was 11;  we therefore calculated each statistic for $k=9$ and 13 and treat half the difference between the mean (across the five folds) of the performance metrics for each $k$ value as the uncertainty on that statistic associated with the choice of $k$. The uncertainties associated with the choices of $k$ value for \umapz\ and the embedding hyperparameters for the SOM and UMAP (sigma, learning rate, \nneighbors) were also calculated in this manner.} 

\added{In order to quantify the uncertainty in each metric due to the finite size of the test used to calculate test statistics we used bootstrap resampling. We created 25 bootstrap realizations of the test set (with replacement) and calculated the standard deviation of each metric across them as an estimate of this uncertainty. To calculate the total uncertainties in a given metric (corresponding to the error bars in Figs. \ref{fig:binned_stat}--\ref{fig:binned_colors} and the errors presented in Table \ref{tab:global}), we add the contributions from each relevant source in quadrature. \figref{uncs_colorsz} shows the contributions to the uncertainty on each performance metric from each uncertainty source, binned by redshift.}

\subsection{\added{Uncertainties in Statistics for \somz\ and \nearestz\ }}
\label{app:somz_unc}

\begin{figure*}
\centering
\includegraphics[width=\textwidth]{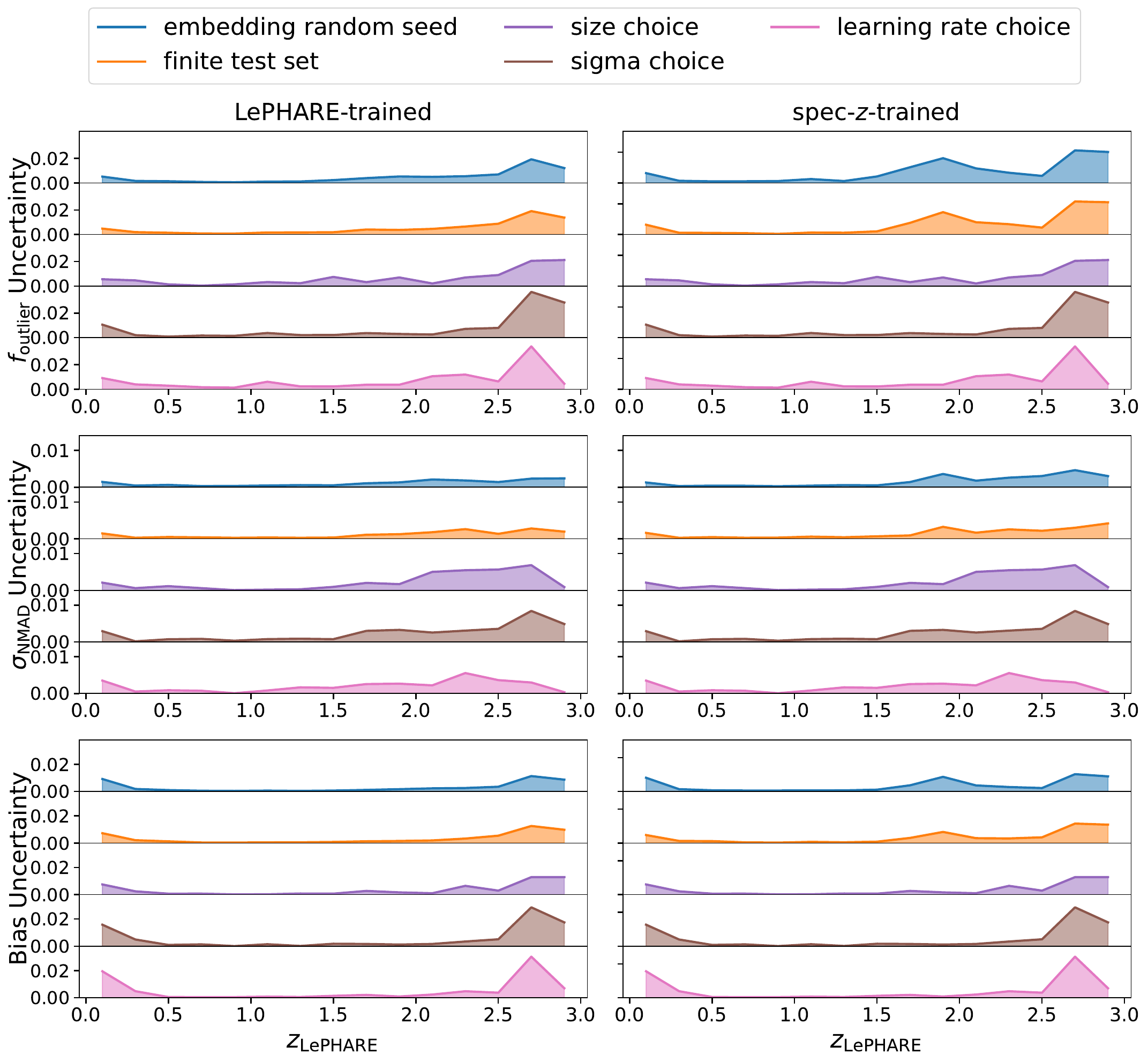}
\caption{\label{fig:uncs_somz}
\added{
Contributions to the uncertainties in the performance metrics of \somz\ predictions  (i.e., those predicted from the median redshift of training objects in a cell, with no prediction for objects in cells without training redshifts; see \secref{interpolation}) as a function of redshift. Note that the sum of the absolute values on the y-axes are not equal to the sizes of the error bars in \figref{binned_stat}, as the latter combines all uncertainties in quadrature. Consistent colors are used for uncertainty sources shared across any of the four redshift prediction methods.}
}
\end{figure*}

\begin{figure*}
\centering
\includegraphics[width=\textwidth]{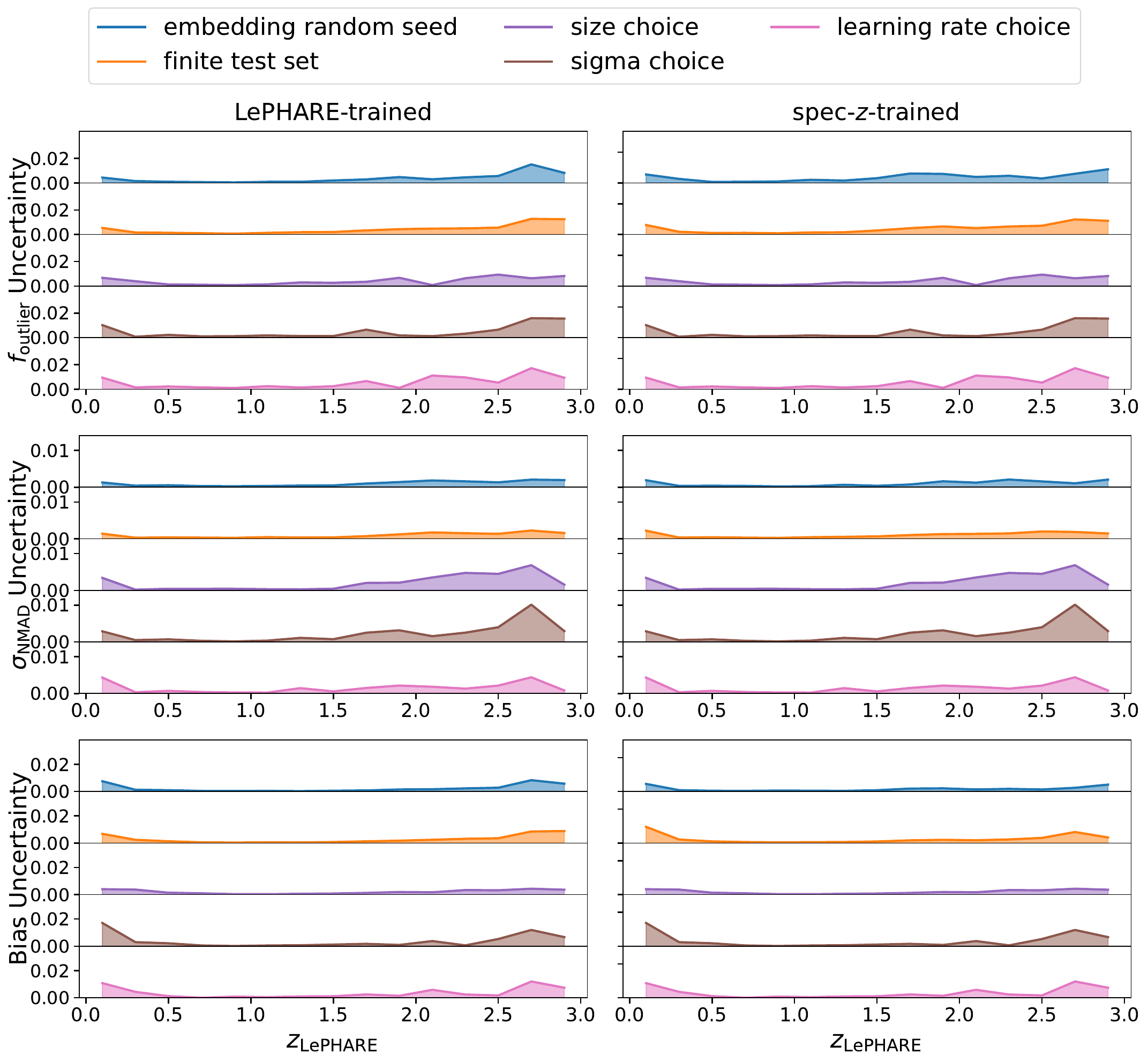}
\caption{\label{fig:uncs_nearestz}
\added{As Fig. \ref{fig:uncs_somz}, but instead showing contributions to the uncertainties in the performance metrics of \nearestz\ predictions  (i.e., those predicted from the median redshift of training objects in a cell, using the nearest available cell if there are no training objects in the original cell; see \secref{interpolation}) as a function of redshift. 
}
}
\end{figure*}

\added{The \somz\ and \nearestz\ algorithms introduced in \secref{interpolation} share the same set of hyperparameters. The uncertainties associated with their values were quantified using the same methods as for the uncertainties associated with the choice of $k$ for \colorsz\ and \umapz\ and the choice of the UMAP hyperparameter \nneighbors. 
For the grid size, to vary the total number of cells by $\pm15\%$ while maintaining an $1\times2$ aspect ratio required grids of $69\times138$ and $80\times160$ cells (as compared to the nominal $75\times150$). For sigma (nominally 1.0) we used values of 0.85 and 1.15, and for learning rate (nominally 0.8) we used values of 0.68 and 0.92.}

\added{The statistics for these algorithms are also affected by uncertainties due to the choice of random seed used in generating the SOM embedding and due to the finite size of the test set. Since these do not involve any hyperparameter choices, we use the full training and testing datasets rather than the validation dataset to calculate these uncertainties (separately for the \lephare-trained and \specz-trained scenarios).}

\added{To determine the uncertainties associated with the choice of random seed, we generated 25 sets of SOM embeddings and redshift predictions with the optimized hyperparameter values and random seed values from 0 to 24, and then calculated the resulting performance metrics. The mean performance metrics across this set of random seed values correspond to the quantities reported in \secref{redshift_estimation} and referenced throughout the paper. The standard deviation of a given statistic across the different random seed values provides our estimate of the uncertainty associated with seeding. }

\added{To quantify the uncertainty due to the finite size of the test set, we performed 25 bootstrap resamplings with replacement of the test samples (all using the same random seed value and SOM embedding), taking the standard deviations of the performance metrics across the resulting samples as the associated uncertainty. 
\figrefs{uncs_somz}{uncs_nearestz} show the contributions from different uncertainty sources to the overall errors on each performance metric for \somz\ and \nearestz, respectively. Generally, the uncertainties for \somz\ are larger than those for \nearestz. We ascribe this to the fact that objects in cells without training data are ignored in the performance metric calculations for \somz, so different subsets of the sample will be used to estimate a given statistic in different random seed or bootstrap realizations, causing a greater degree of variation. In the case of the uncertainties associated with hyperparameter choices, this phenomenon is somewhat exacerbated by the limited number of objects in the validation set, especially at high-$z$.}

\subsection{\added{Uncertainties in Statistics for \umapz}}
\label{app:umapz_unc}

\begin{figure*}
\centering
\includegraphics[width=\textwidth]{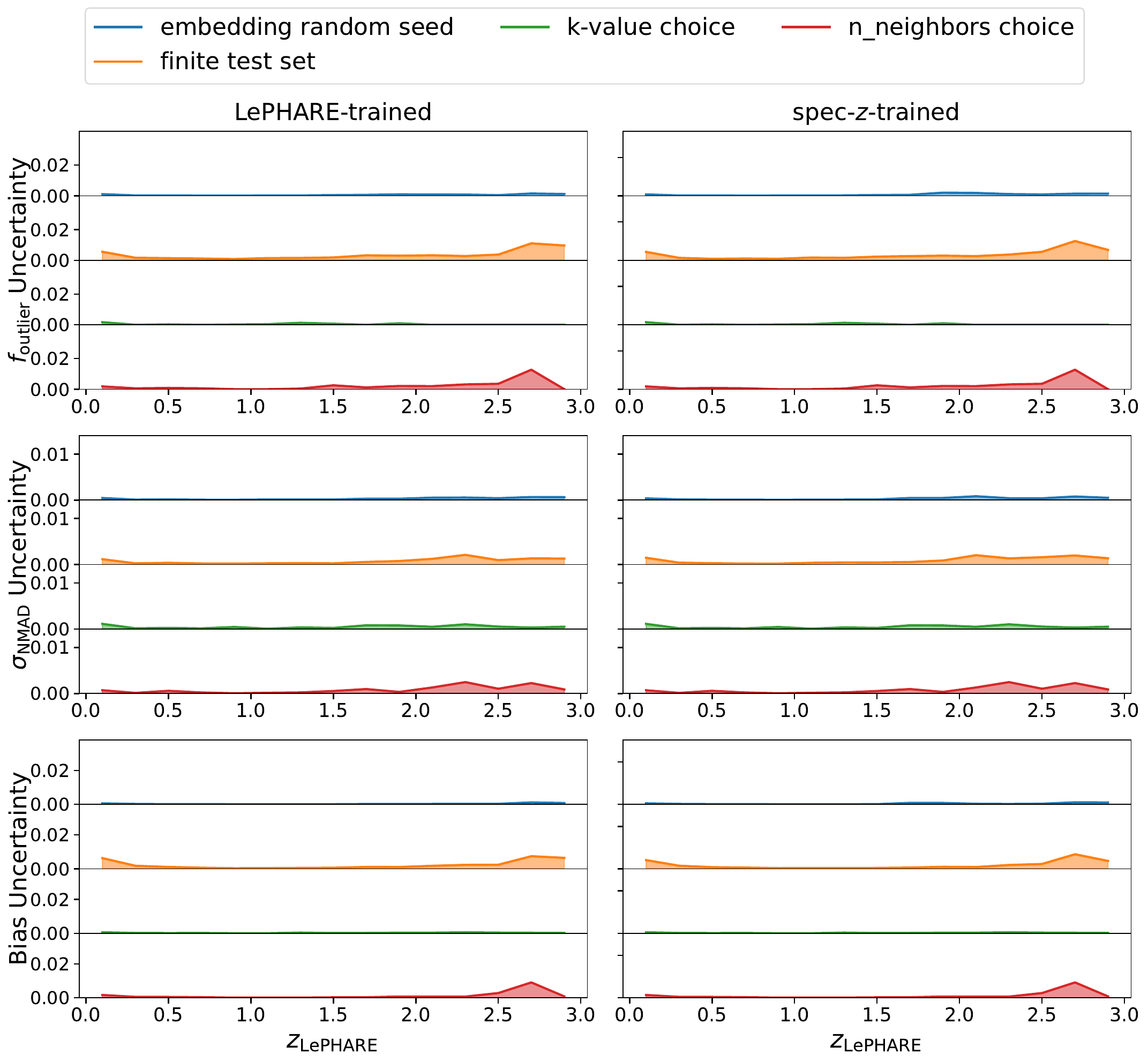}
\caption{\label{fig:uncs_umapz}
\added{As Fig. \ref{fig:uncs_somz}, but instead showing contributions to the uncertainties in the performance metrics for \umapz\ predictions  (i.e., those predicted from the nearest neighbors in UMAP coordinates; see \secref{interpolation}) as a function of redshift.
Consistent colors are used for uncertainty sources shared across any of the four redshift predictors.}
}
\end{figure*}

\added{For \umapz\ (described in \secref{interpolation}), the hyperparameters that could contribute to uncertainties included the choice of the \nneighbors\ value used in generating the embeddings and the choice of $k$ value for the redshift estimator. As in Sections \ref{app:colorsz_unc} and \ref{app:somz_unc}, we used five-fold cross-validation using the validation set and varied parameters by $\pm15\%$ to quantify uncertainties associated with hyperparameter choices. For \nneighbors\ (nominally 70) we calculated statistics for values of 69, and 80, and for $k$ (nominally 15) we used values of 13 and 17.}

\added{The \umapz\ performance statistics will also be affected by uncertainties associated with the choice of the random seed used in generating the embedding and by the finite size of the test set. 
To assess the impact of the random seeds, we generated 25 UMAP embeddings with the optimized hyperparameters and random seeds 0-24 and estimated the redshifts of the testing samples using \umapz\, similarly to what was done for \somz\ and \nearestz. The mean value of a given performance metric amongst these embeddings corresponds to the metric values presented throughout the paper and in \figref{binned_stat}; the uncertainty in a given metric due to variation in random seeds is then taken as its standard deviation across the 25 embeddings.}

\added{As before, the uncertainty in a given statistic due to the finite size of the test set is quantified by its standard deviations over 25 bootstrap resamplings of the test samples with redshift estimates derived from a fixed UMAP embedding. To calculate total uncertainties (e.g., as presented in Table \ref{tab:global}), all sources of errors are added in quadrature. \figref{uncs_umapz} shows the contributions to the uncertainty on each performance metric from each uncertainty source, binned by redshift. The uncertainties associated with \umapz\ hyperparameter choices are generally smaller than those for the SOM-based methods. In addition, the scatter in performance statistics associated with variations in the random seeds used to generate an embedding are much smaller for \umapz\ than for \somz\ or \nearestz.}

\subsection{\added{Comparison of Combined Uncertainties in Performance Metrics}}
\label{app:compare_uncs}

\begin{figure*}
\centering
\includegraphics[width=\textwidth]{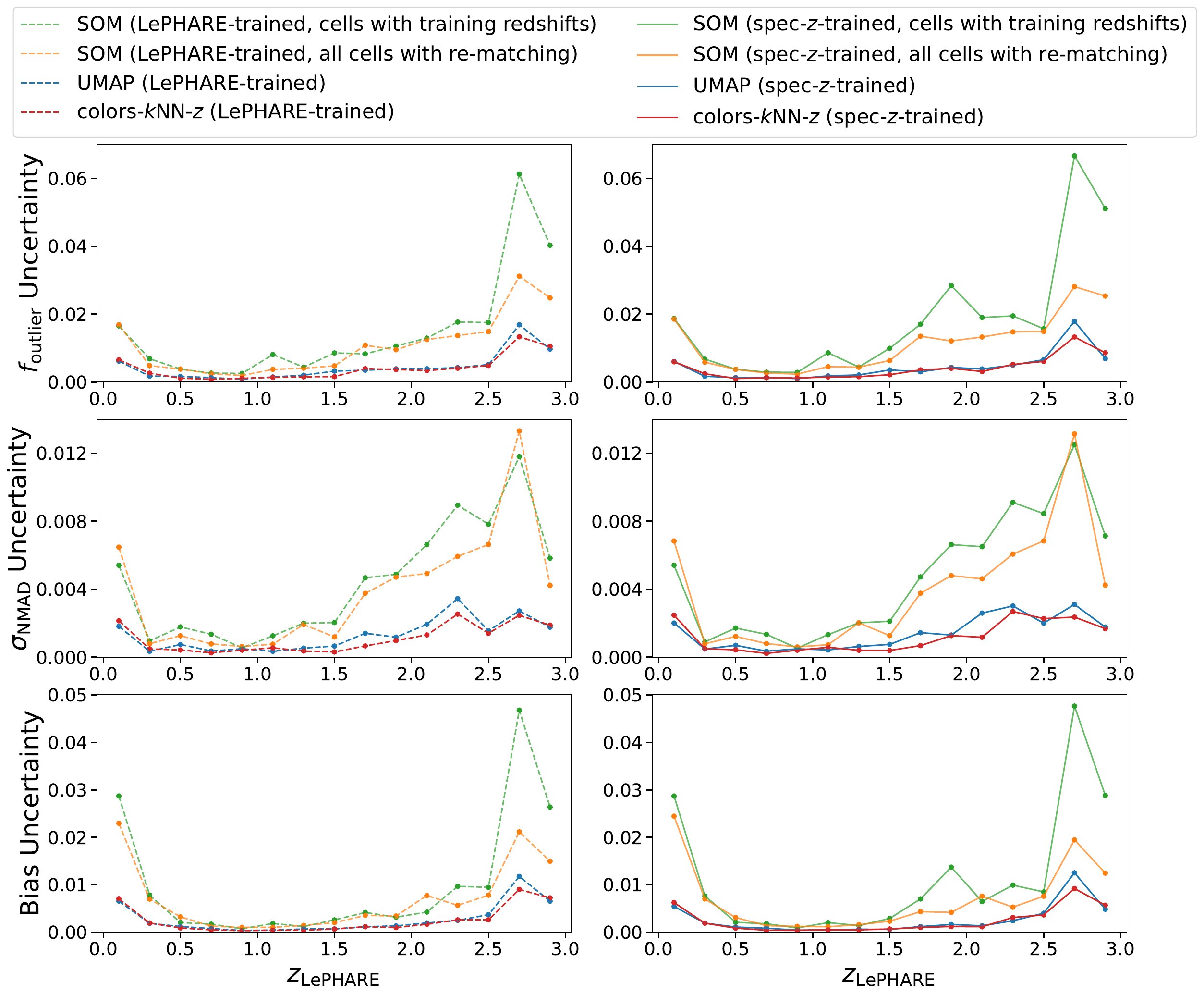}
\caption{\label{fig:compare_uncs}
\added{Total uncertainties (added all contributions in quadrature) for each prediction method: SOM using only the objects in cells with training redshifts (\somz, green) or re-matching the objects in cells without training data to the nearest available cells (\nearestz, orange); $k$NN interpolation in the $ugrizyJH$ color space (\colorsz, red); and $k$NN interpolation in the UMAP space (\umapz, blue). The values shown here are equivalent to the sizes of the error bars in \figrefs{binned_stat}{binned_colors}, and to the quadrature sums of the uncertainties plotted in Figures \ref{fig:uncs_colorsz} through \ref{fig:uncs_umapz}. The uncertainties for both SOM-based techniques are much larger than those for \colorsz\ or \umapz, particularly in the more sparsely-sampled regimes at $z<0.5$ and $z>1.5$ (see \figref{non_IID}). 
The similarity between errors for \colorsz\ and \umapz, by contrast, indicates that although new potential sources of variability in performance (namely, the random seed choice and \nneighbors\ value) are introduced by the UMAP dimensionality reduction step, these have little impact on the reproducibility of our metrics.
}
}
\end{figure*}

\added{In \figref{compare_uncs}, we show the total uncertainties (adding all contributions in quadrature) for each of the four redshift prediction algorithms considered in this paper. The plotted values for each bin and method are equivalent to the sizes of the error bars in \figrefs{binned_stat}{binned_colors}. 
By comparing the total uncertainties for all four methods directly in this way, 
we can gain some insight into how sensitive the redshift predictions from each method are to the details of the low-dimensional SOM and UMAP embeddings and the hyperparameters used to construct them.
}

\added{In particular, the scatter in performance metrics for the SOM-based techniques are generally much greater than those for \colorsz\ or \umapz. This is true for all three performance metrics, and the effect is most severe in the sparsely-sampled redshift regimes at $z<0.5$ and $z>1.5$ (see \figref{non_IID}). In the more densely-sampled regime between these extremes, the uncertainties are more similar among all four methods, though the SOM-based approaches still yield somewhat higher uncertainties than \colorsz\ and \umapz\ for \fout\ and \nmad\ (and marginally higher uncertainty on bias). This parallels the trends for the values of the performance metrics themselves, as seen in \figrefs{binned_stat}{binned_colors}, where all four methods yielded broadly similar performance in the $0.5<z<1.5$ regime. 
}

\added{In contrast to the behavior of the SOM-based approaches, the uncertainties for \umapz\ are similar to those for \colorsz\ across the full redshift range, despite the UMAP approach introducing two additional sources of variation associated with the dimensionality reduction step (namely, the embedding random seed and the choice of \nneighbors\ value). Taken together with the similar values of the performance metrics themselves for the two methods, 
this suggests that the UMAP embeddings have preserved the color information most critical to redshift estimation in a highly reproducible manner. This is in juxtaposition to the SOM-based approaches, which yield much larger errors at redshifts outside of $0.5<z<1.5$ due primarily to larger uncertainties associated with dimensionality reduction. As seen in \figref{uncs_umapz}, the embedding random seed does not contribute significantly to the uncertainties on \umapz\ performance, while \figrefs{uncs_somz}{uncs_nearestz} show that it does contribute notably to the uncertainties on the SOM-based methods, especially at the most sparsely-sampled redshifts. The same figures show that the SOM hyperparameters (grid size, sigma, and learning rate) generally cause larger variations in performance than the UMAP hyperparameters do. 
The other uncertainty source unique to \umapz\ (with respect to the SOM methods) is the $k$ value, but it contributes only marginally to variations in performance.
}

\added{
Overall, comparing the uncertainties computed for the four redshift prediction methods shows that the compressed color spaces created with UMAP lead to highly reproducible results with minimal uncertainty associated with random seed variations, and are furthermore insensitive to shifts of up to $\pm15\%$ in the value of the key embedding hyperparameter choice, \nneighbors. Performing the analogous uncertainty quantification procedure for SOM-based approaches, we find that the variability associated with variations in the random seed or small shifts in the hyperparameter values is much greater.}

\added{Although we have focused on quantitative assessment of the impact of these variations on the ability to predict redshift from position in a dimensionality-reduced space, we note that the greater stability of the UMAP embedding is also apparent when visualizing the dimensionality-reduced spaces directly. Changing random seeds or hyperparameters yields UMAP embeddings with very similar structure and mapping to redshift.  In contrast, even with fixed hyperparameters SOM maps with different random seeds appear very different, with the density of objects in cells and their median redshifts being very different from one embedding to another.  These qualitative variations clearly contribute to the quantitative variability explored in this Appendix.
}

\bibliography{library}{}
\bibliographystyle{aasjournal}

\end{document}